\documentclass[aps,prd,floats,preprintnumbers,nofootinbib]{revtex4}
\usepackage{epsfig}

\newcommand{\lesim}{\,\raisebox{-.3ex}{$_{\textstyle <}\atop^{\textstyle\sim}$}\,}

\begin{document}

\preprint{FERMILAB-Pub-03/206-T}
\preprint{nuhep-exp/2003-01}

\title{Invisible $Z$-Boson Decays at $e^+e^-$ Colliders}
 
\author{Marcela Carena}
\affiliation{Theoretical Physics Division, Fermilab, P.O. Box 500, Batavia, IL~60510, USA}
 
\author{Andr\'e de Gouv\^ea}
\affiliation{Theoretical Physics Division, Fermilab, P.O. Box 500, Batavia, IL~60510, USA}

\author{Ayres Freitas}
\affiliation{Theoretical Physics Division, Fermilab, P.O. Box 500, Batavia, IL~60510, USA}

\author{Michael Schmitt}
\affiliation{Northwestern University, Department of Physics \& Astronomy, 
2145 Sheridan Road, Evanston, IL~60208, USA}

\begin{abstract}
The measurement of the invisible $Z$-boson decay width at $e^+ e^-$ 
colliders can  be done ``indirectly,'' by subtracting the $Z$-boson visible partial 
widths from the $Z$-boson total  width, or ``directly,'' from 
the process $e^+ e^- \rightarrow \gamma \nu 
\bar{\nu} $.  Both procedures are sensitive to different types of new 
physics and provide information about the couplings of the neutrinos to the $Z$-boson.  
At present, measurements at LEP and CHARM~II are capable of constraining the 
left-handed $Z \nu \bar{\nu}$-coupling, $ 0.45 \lesim g_L \lesim 0.5 $, while the 
right-handed one is only mildly bounded, $|g_R| \leq 0.2$. We 
show that measurements at a future $e^+ e^-$ linear collider 
at different center-of-mass energies, $\sqrt{s} = m_Z$ and 
$\sqrt{s} \approx 170$~GeV, would translate into a 
markedly more precise measurement of the $Z\nu \bar{\nu}$-couplings. A 
statistically significant deviation from Standard Model predictions
will point toward different new physics mechanisms, depending on whether
the discrepancy appears in the direct or the indirect measurement of the 
invisible $Z$-width. We discuss some scenarios which illustrate the ability of
different invisible $Z$-boson decay measurements to constrain new 
physics beyond the Standard Model.
\end{abstract}
 
\maketitle

\setcounter{equation}{0}
\section{Introduction}

The four LEP collider experiments have performed several precise
measurements of the properties of the $Z$-boson 
\cite{LEP_Z,PDG,combined,combined_2}, the heavy,
neutral partner of the $W$-boson and the photon. These measurements are
part of the evidence that the Standard Model of the Electroweak Interactions (SM)
works extremely well, up to energies of several hundred~GeV. 
One of these measurements is associated with the ``invisible $Z$-boson width'' 
(invisible $Z$-width). Assuming that the SM is correct, this measurement 
can be translated into a count of the number of neutrino species. The current value of 
the invisible $Z$-width agrees quite well with the SM 
expectation that there are three very light ($m_{\nu}\ll 1~\rm GeV$) 
neutrino species. This is often interpreted as evidence that the SM contains 
three and only three families of fermionic fields, meaning that there is no 
fourth sequential generation. It is remarkable that this result is in agreement with
cosmological constraints on the number of relativistic species around the time of 
Big Bang nucleosynthesis, which seem to indicate the existence of three very light 
neutrino species \cite{BBN}. 

It is interesting to note that the LEP result is precise enough
to probe whether the ``number of neutrinos,'' $N_{\nu}$,
deviates slightly from three. Indeed, it is often quoted that the most 
precise LEP numbers can be translated into $N_{\nu}=2.9841\pm0.0083$ \cite{PDG}, about 
two sigma away from the SM expectation, $N_{\nu}=3$. 
While not statistically significant, this result has invited 
theoretical speculations, some of which involve suppressing the 
$Z\nu\bar{\nu}$-couplings with respect to the SM value. 

More recently, the NuTeV Collaboration presented a measurement of
$\sin^2\theta_W$ obtained from neutrino--nucleon scattering \cite{NuTeV}. This result 
overshoots the SM prediction by about three sigma ($\sin^2\theta_{W}=0.2277\pm0.0016$~(NuTeV) 
versus $\sin^2\theta_{W}=0.2227\pm0.0004$ (SM prediction), see \cite{NuTeV}). 
One possible explanation of this ``NuTeV anomaly'' is that the $Z\nu\bar{\nu}$-couplings are 
suppressed (by a factor $\rho_0=0.9941\pm0.0021$ \cite{NuTeV2}) 
with respect to their SM values \cite{NuTeV,NuTeV2,theory_nutev,NuTeV_review}. 

In light of these two either statistically weak (the invisible $Z$-width at LEP) or controversial (the 
NuTeV anomaly) discrepancies, the particle physics community would profit
from other independent, precise measurements of the $Z\nu\bar{\nu}$-couplings.
We argue that a linear collider experiment, capable of taking data around and above the 
$Z$-boson mass, can provide useful, precise, and, more importantly, ``different''
measurements with invisible $Z$-boson decays. 

One reason for this is that the most precise LEP measurement of the invisible 
$Z$-width is {\sl indirect}, in the sense that it is determined by subtracting the 
$Z$-boson visible partial widths from the $Z$-boson total width. We argue that at a 
(much) higher statistics linear collider experiment a competitive, {\sl direct} 
measurement of the invisible $Z$-boson width can be obtained from the process 
$e^+e^-\to\gamma\nu\bar{\nu}$ by counting events with a photon plus missing energy.

The indirect and direct measurements of the
invisible $Z$-width are sensitive to different types of
physics beyond the SM in different ways. While in some scenarios ({\it e.g.}\/
modified $Z\nu\bar{\nu}$-couplings) the two results should be identical (and, perhaps,
different from SM expectations), in other scenarios ({\it e.g.}\/ a non-zero 
$\gamma\nu\bar{\nu}$-coupling) the two measurements of the invisible $Z$-width may disagree.  

Furthermore, the very precise LEP result, obtained at center-of-mass energies around
the $Z$-boson mass, is, in practice, only sensitive to a particular combination
of the $Z\nu\bar{\nu}$-couplings: $g_L^2+g_R^2$, where $g_L$ ($g_R$) is the 
left-handed (right-handed) $Z\nu\bar{\nu}$-coupling. In the SM, the neutrinos only couple
left-handedly to the $Z$ and $W$-bosons, but the 
``left-handedness'' of the $Z\nu\bar{\nu}$-couplings has not been experimentally 
established with good precision. Some information on $g_L$ and $g_R$ can be
obtained, under a specific set of assumptions, by combining the LEP result with results 
from neutrino--electron scattering. Furthermore, by looking
at $e^+e^-\rightarrow\gamma\nu\bar{\nu}$ at center-of-mass energies above the 
$Z$-boson mass, one is sensitive to both $(g_L^2+g_R^2)$ and 
$g_L$ separately, thanks to the interference between the $s$-channel $Z$-boson exchange
and the $t$-channel $W$-boson exchange. This means that by analyzing LEP data at
center-of-mass energies above the $Z$-boson mass one can also learn about the individual
values of $g_L$ and $g_R$. A linear collider experiment taking data above the $Z$-boson mass 
(at, for example, $\sqrt{s}=170$~GeV) can perform a more precise 
(and less model dependent) measurement of $g_L$ and $g_R$, as will be studied in detail.  

This manuscript is organized as follows.
In Sec.~\ref{section_Zpole}, we discuss in some detail 
the LEP measurements of the invisible $Z$-width, emphasizing the assumptions that
are made in order to obtain the precise value of $N_{\nu}$ quoted above. Having done
that, we discuss how precisely one should be able to directly measure the 
invisible $Z$-width at a linear collider operating at center-of-mass energies
``around'' the $Z$-boson mass. In Sec.~\ref{section_away}, we discuss how
one should be able to measure the neutrino $g_L$ and $g_R$ couplings separately by
taking $e^+e^-$ data at center-of-mass energies higher than the $Z$-boson mass.
We look at current constraints that can be obtained from combining LEP data with data on
neutrino--electron scattering, and then examine the existing LEP data collected above 
the $Z$-boson mass (LEP II). We proceed to discuss how well a similar procedure
can be executed at a linear collider. In Sec.~\ref{section_theory}, we analyze new
physics contributions that would lead to discrepancies between the SM and the 
``measurements'' which are proposed above. A summary of the results and some parting
thoughts are presented in Sec.~\ref{section_conclusions}.

\setcounter{equation}{0}
\section{The Invisible $Z$-boson Width Around the $Z$-pole}
\label{section_Zpole}

The SM predicts that around $20$\% of the time a $Z$-boson will decay
into a $\nu\bar{\nu}$ pair. The neutrino pair cannot be observed directly in
collider experiments, meaning that $Z$-bosons decaying in this fashion
are ``invisible.''

In electron--positron colliders there are 
two ways of establishing
whether these invisible $Z$-boson decays are occurring, and to measure the
invisible $Z$-width. 
One is to directly measure the total $Z$-width, $\Gamma_{\rm tot}$,
by studying the line-shape of the $Z$-boson (this is done by colliding $e^+e^-$ 
at center-of-mass energies around the $Z$-boson mass), 
and measuring its partial decay widths in visible final states, 
$\Gamma_{\rm vis}$, namely charged leptons and hadrons. One can then compute 
the invisible $Z$-width, $\Gamma_{\rm inv}$: 
$\Gamma_{\rm inv}=\Gamma_{\rm tot}-\Gamma_{\rm vis}$. This procedure is discussed
in detail below, in Subsection \ref{section_Zpole}A.
The other is to look for events where an initial state lepton
radiates off a hard photon before annihilating into an $s$-channel $Z$-boson.
When that happens, if the $Z$-boson decays invisibly, the experimentally 
observed final state is a single photon plus a significant amount of 
``missing energy'' (in summary, $e^+e^-\rightarrow \gamma Z
\rightarrow\gamma\nu\bar{\nu}$). This procedure will be discussed in detail
in Subsection \ref{section_Zpole}B.

\subsection{On the LEP (Indirect) Measurement of the Invisible $Z$-boson Width}

At LEP, the invisible $Z$-width is indirectly extracted from the following
observables:
\begin{itemize}
\item{ 
$\Gamma_{\rm tot}=2.4952\pm0.0023$~GeV, the total width of the 
$Z$-boson\footnote{These are the combined values obtained by the LEP Electroweak
Working Group \cite{combined,combined_2}.} and $m_Z=91.1876\pm0.0021$~GeV, the $Z$-boson pole mass;}

\item{
$\sigma_h^0=41.541\pm0.037$~nb, the hadronic pole cross section, defined as
\begin{equation}
\sigma_h^0\equiv \frac{12\pi}{m_Z^2}\frac{\Gamma_{ee}\Gamma_{\rm had}}
{\Gamma_{\rm tot}^2},
\end{equation}
where $\Gamma_{ee}$ and $\Gamma_{\rm had}$ are the partial $Z$-boson decay
widths into an $e^+e^-$ pair and into hadrons respectively;}

\item{
$R_{\ell}=20.804\pm0.050,~20.785\pm0.033,~20.764\pm0.045$ for 
$\ell=e,\mu,\tau$, respectively, defined as $\Gamma_{\rm had}/
\Gamma_{\ell\ell}$. If one assumes universal $Z$-boson
couplings to charged leptons, $R_{\ell}=20.767\pm0.025$.}
\end{itemize}

Assuming lepton universality and taking into account the fact that several
of the measurements listed above are strongly correlated, one can easily
compute the invisible $Z$-width
and obtain the LEP result, which is quoted
by the Particle Data Group (PDG) \cite{PDG}:
\begin{equation} 
\Gamma_{\rm inv}^{LEP}=499.0\pm1.5~{\rm MeV}.
\label{LEP_res}
\end{equation} 
This result is to be compared to the SM prediction,
\begin{equation} 
\Gamma_{\rm inv}^{SM}=501.3\pm0.6~{\rm MeV}, 
\label{invZ_SM}
\end{equation}
meaning that 
$\Delta\Gamma_{\rm inv}\equiv\Gamma_{\rm inv}^{SM}-\Gamma_{\rm inv}^{LEP}=
-2.2\pm1.6$~MeV, a $1.4\sigma$ effect. This result can also be expressed as
an upper bound on additional contributions to the invisible $Z$-width. Numerically,
one obtains $\Gamma^{\rm new}_{\rm inv}<2.0$~MeV at the 95\% confidence level,
assuming that the new physics contributions adds incoherently with the neutrino
pair production ({\it i.e.,}\/ $\Gamma^{\rm new}_{\rm inv}$ is strictly positive).

In order to obtain the well known $2\sigma$ discrepant measurement of the number
of neutrinos, one should consider the ratio of partial widths
\begin{equation}
\frac{\Gamma_{\rm inv}}{\Gamma_{\ell\ell}}\equiv N_{\nu}\left(\frac{\Gamma_{\nu\nu}}
{\Gamma_{\ell\ell}}\right)_{SM}.
\label{number_of_nus}
\end{equation}
Eq.~(\ref{number_of_nus}) defines what is meant by the ``number of
neutrinos.'' $N_{\nu}$ only agrees with the {\it de facto} number of neutrinos if both the
$Z\ell\bar{\ell}$ and the $Z\nu\bar{\nu}$-couplings have their SM predicted values.  
The SM prediction for $(\Gamma_{\nu\nu}/
\Gamma_{\ell\ell})_{SM}=1.9912\pm0.0012$ is more precisely known than the 
individual partial widths, and when compared to the extracted value of 
$\Gamma_{\rm inv}/\Gamma_{\ell\ell}=5.942\pm0.016$ yields 
\begin{equation}
N_{\nu}^{\rm LEP}=2.9841\pm0.0083, 
\label{N_nus_LEP}
\end{equation}
the result we alluded to in the Introduction.

The results Eq.~(\ref{LEP_res}) and Eq.~(\ref{N_nus_LEP})
imply different consequences for different SM extensions. For example, modified
$Z\nu\bar{\nu}$-couplings combined with {\sl identically} 
modified $Z\ell^+\ell^-$-couplings would ideally
lead to a nonzero $\Delta\Gamma_{\rm inv}$ 
but to a zero $N_{\nu}-3$. Furthermore, given the indirect way that $\Gamma_{\rm inv}$ is
extracted, one should be careful when it comes to defining what 
$\Delta\Gamma_{\rm inv}$ is really sensitive to. The observation of a 
discrepant $\Gamma_{\rm inv}$ and/or $N_{\nu}$, does {\sl not} necessarily
imply that there is new physics in the neutrino sector or even in the leptonic sector.
For example, it is possible that other effects may modify the extracted value of 
$\Gamma_{\rm tot}$, hence inducing a discrepancy between the measured 
invisible $Z$-width and its SM prediction. This will be further explored in 
Sec.~\ref{section_theory}.

We have also extracted the value of the invisible $Z$-width without
assuming lepton universality
 and, using the
results presented in \cite{combined,combined_2}, obtained
\begin{equation}
\Gamma_{\rm inv}^{LEP}{\rm (nonuniversal)}=497.4\pm2.5~{\rm MeV},
\end{equation}
less precise than the result obtained assuming universality, as expected. 
In spite of that, $\Delta\Gamma_{\rm inv}$(nonuniversal)$=-3.9\pm2.6$~MeV, 
still a $1.5\sigma$ deviation, is as significant as the effect obtained 
assuming universality. This result translates into an upper
bound on $\Gamma^{\rm new}_{\rm inv}<3.2$~MeV at the 95\% confidence level,
assuming that the new physics effect does not interfere with the neutrino--antineutrino
final state.

An attempt to extract the number of neutrinos via
Eq.~(\ref{number_of_nus}) without charged-lepton universality would be 
rather peculiar, since one needs to explicitly assume that 
$\Gamma_{\nu_e\nu_e}=\Gamma_{\nu_{\mu}\nu_{\mu}}=
\Gamma_{\nu_{\tau}\nu_{\tau}}$ in order to relate $N_{\nu}$ to a ``neutrino
number.'' Nonetheless, one
can easily extract the value of $\Gamma_{\rm inv}/\Gamma_{ee}$ and 
$\Gamma_{\rm inv}/\Gamma_{\mu\mu}$, and compute, respectively, $N_{\nu}^{ee}$ and
$N_{\nu}^{\mu\mu}$, these being defined via Eq.~(\ref{number_of_nus}) with 
$\Gamma_{\ell\ell}$ replaced, respectively, by $\Gamma_{ee}$ and $\Gamma_{\mu\mu}$.
We obtain
\begin{eqnarray}
N_{\nu}^{ee}&=&2.978\pm0.012, \\
N_{\nu}^{\mu\mu}&=&2.973\pm0.019.
\end{eqnarray}
These are, respectively, 1.8$\sigma$ and 1.4$\sigma$ away from the SM prediction of
$N_{\nu}=3$.

\subsection{(Direct) Measurement of the Invisible $Z$-boson Width at a Linear 
Collider}

\begin{figure}[t]
{\centering
\epsfig{figure=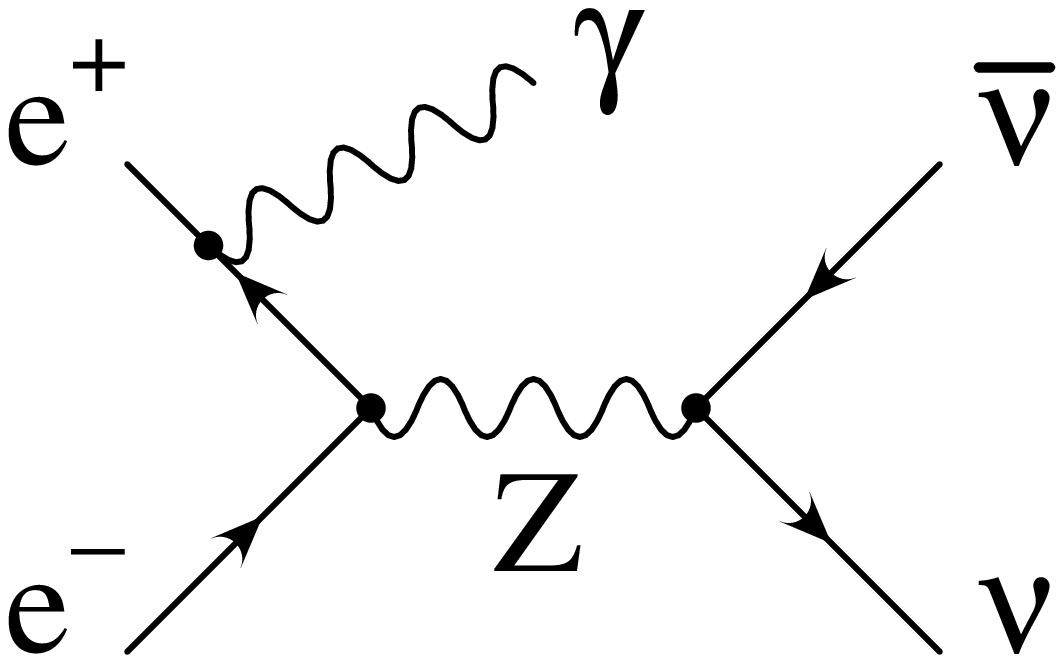,width=4.5cm}
\epsfig{figure=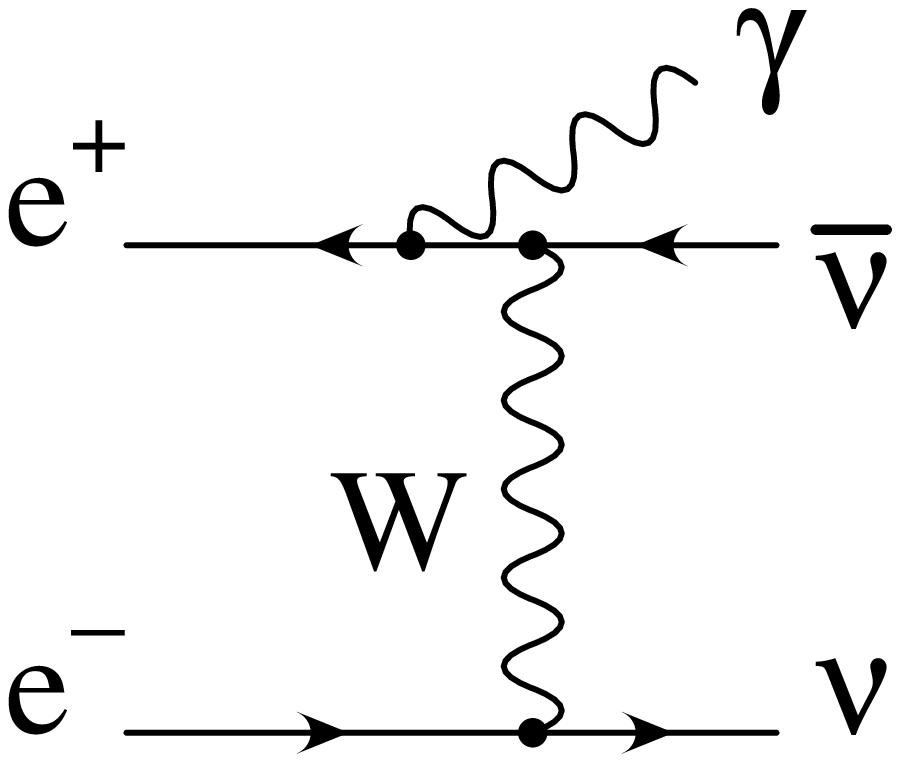,width=4.5cm}
\epsfig{figure=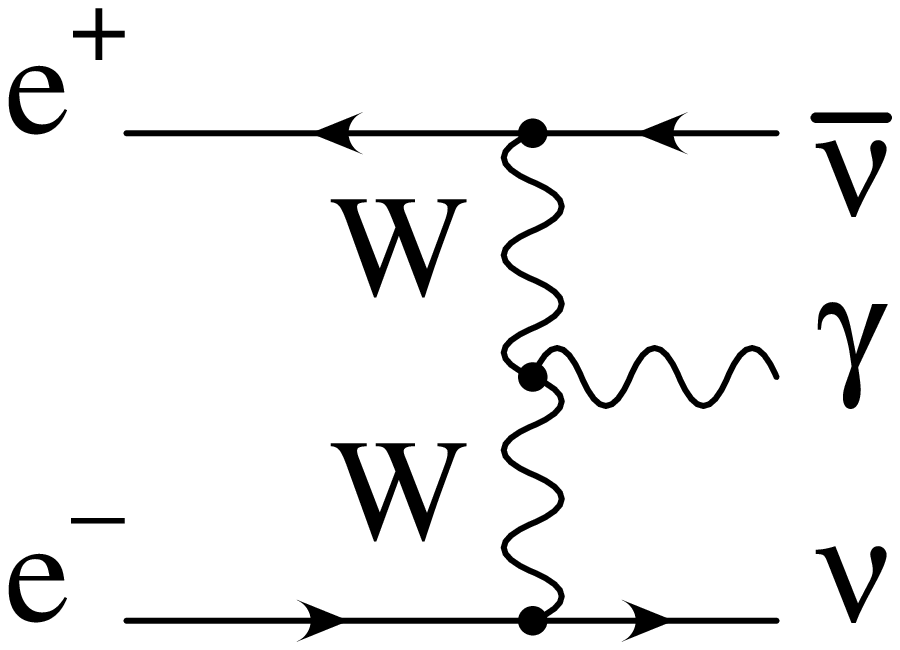,width=4.5cm}
}
\caption{Leading order Feynman diagrams contributing to $e^+e^-\rightarrow\gamma\nu\bar{\nu}$.
\label{fig_diagrams}
}
\end{figure}

In the SM, for center-of-mass energies around the $Z$-boson
mass, the dominant contribution to $e^+e^-\to\gamma+$ missing energy 
comes from an intermediate
$Z\gamma$ pair, followed by $Z\rightarrow\nu\bar{\nu}$. Other contributions
come from $t$-channel $W$-boson exchange, plus one photon vertex attached either
to the initial state electrons or to the intermediate state charged gauge boson. The leading
order Feynman diagrams are depicted in Fig.~\ref{fig_diagrams}.

The LEP collaborations have measured the cross section for the photon plus missing
energy final state. The most precise
result comes from the L3 experiment, after analyzing of 100~pb$^{-1}$ of data: 
$\Gamma_{\rm inv}=498\pm12\pm12$~MeV \cite{L3_direct}
(the first error is due to statistics, while the second one to systematics). 
Other LEP collaborations have older results
\cite{ALEPH_direct,OPAL_direct,DELPHI_direct,others_direct} 
(from smaller data samples) with errors, both statistical
and systematic, which are two to six times larger. 
The PDG average is $\Gamma_{\rm inv}=503\pm16$~MeV \cite{PDG}, 
dominated by the L3 result \cite{L3_direct}. For later comparison, it proves useful to 
estimate the ultimate sensitivity for LEP (including all four experiments
analyzing all the data collected around the $Z$-boson mass). We do this by naively rescaling
the L3 result from 100~pb$^{-1}$ to roughly 500~pb$^{-1}$. Assuming that both the
statistical error and the systematic error will decrease by a factor $\sqrt{500/100}$,
we obtain $\delta\Gamma_{\rm inv}=\pm5\pm5$~MeV. 

The relatively large (compared to the indirect result) 
error of the direct measurement of $\Gamma_{\rm inv}$
reflects the small statistical sample of $e^+e^- \to \gamma \nu \bar{\nu}$ 
events available at LEP. Therefore, a significant
improvement can be expected from a high-luminosity linear collider running around
the $Z$-boson mass.
At such a ``Giga-Z'' machine, it is envisaged that within 100 days of running,
a sample of $10^9$ $Z$-boson decays can be collected~\cite{LC}.

Assuming 50~fb$^{-1}$ of $e^+e^-$ data collected around the $Z$-boson mass, we examine
how this would improve the measurement of the invisible $Z$-width.
We are mostly
interested in the result that could be obtained by looking for $\gamma +$
missing energy, but will first briefly present the improvement that can
be expected for determining the invisible $Z$-width indirectly, as discussed
in the previous subsection. We assume \cite{TESLA_Zparam,TESLA_study} 
that the total $Z$-width can be measured a factor of roughly two times more 
precisely, $\delta\Gamma_{\rm tot}=\pm 1$~MeV, while $R_{\ell}$ (assuming 
universality) and $\sigma_h^0$ will be measured with uncertainties 
$\delta(R_{\ell})=\pm0.018$ and $\delta(\sigma_h^0)=\pm0.03$~nb (most conservative 
scenario), or $\delta(R_{\ell})=\pm0.004$ and $\delta(\sigma_h^0)=\pm0.015$~nb 
(most optimistic scenario). We refer readers to \cite{TESLA_Zparam,TESLA_study} 
for more details.
Further assuming that the correlation matrix between the observables is
identical to the one obtained for the combined LEP results,\footnote{It is likely
that this correlation matrix will be different for the Giga-Z data.  As 
the correlations depend on the details of the analyses, it is not possible for us
to predict them at this time.  Nonetheless, 
our estimates of the uncertainties which can be obtained at a linear collider should be 
trustworthy.} 
we estimate that the invisible $Z$-width can be measured with an uncertainty 
$\delta(\Gamma_{\rm inv})=\pm1.1$~MeV (most conservative) or 
$\delta(\Gamma_{\rm inv})=\pm0.5$~MeV (most optimistic). For the most
optimistic case, the experimental error would be slightly better than the current 
theoretical error for computing the invisible $Z$-width within the SM, 
Eq.~(\ref{invZ_SM}). Compared with the current LEP precision, Eq.~(\ref{LEP_res}), we
therefore expect between a factor 1.3 and a factor 3 reduction of the error on
$\Gamma_{\rm inv}$ from the Giga-Z experiment.

For illustrative purposes, if one assumes that the results for 
$\Gamma_{\rm inv}$ and $\Gamma_{\ell\ell}$ at the Giga-Z 
agree with the central values obtained at LEP, one would measure
$(\Gamma_{\rm inv}/\Gamma_{\ell\ell})^{\rm Giga-Z}=5.942\pm0.012$ (most conservative)
or $(\Gamma_{\rm inv}/\Gamma_{\ell\ell})^{\rm Giga-Z}=5.942\pm0.006$ (most optimistic).
Translating into a ``number of neutrinos'' we would have
\begin{eqnarray}
N_{\nu}^{\rm Giga-Z}&=&2.984\pm0.006~{\rm (most~conservative)}, \\
N_{\nu}^{\rm Giga-Z}&=&2.984\pm0.003~{\rm (most~optimistic)},
\end{eqnarray}
either 2.5$\sigma$ or 5$\sigma$ away from the SM prediction. In the most optimistic case, 
the experimental error would start to approach 
the theoretical error which goes into computing the neutrino to charged-lepton partial 
decay width ratio.\footnote{This theoretical error, however, is currently dominated by 
the uncertainty on the top-quark mass and the uncertainty on the Higgs mass. 
It is likely that by the time a Giga-Z experiment takes data, these two quantities will 
be much better known.} Therefore, given the assumptions outlined
above, the weak 2$\sigma$ effect observed at LEP could grow to something
between a 3$\sigma$ evidence and a 5$\sigma$ discovery that something is ``wrong''
(assuming that this discrepancy is genuine and not just a statistical fluctuation).

A much more significant improvement can be anticipated for the direct measurement of
the invisible $Z$-width. In order to compute the precision to which it 
can be directly measured at the Giga-Z experiment, we calculate the cross section for 
$e^+e^-\rightarrow\gamma+{\rm invisible}$ for different center-of-mass energies. 
We require the photon energy to be above $E_{\gamma}^{\rm min}=1$~GeV, 
and that it is emitted at an angle with respect to the beam axis larger than
$\theta^{\rm min}_{\gamma}=20^\circ$. For any set of cuts on the photon energy and
emission angle, the SM cross section for 
$\nu\bar{\nu}\gamma$ is given by \cite{sigma_gnn}
\begin{equation}
\sigma_{\nu\bar{\nu}\gamma} = \int_{x^{\rm min}}^{1}{\rm d}x
\int^{\cos\theta_{\gamma}^{\rm min}}_{-\cos\theta_{\gamma}^{\rm min}}
{\rm d}y \frac{\alpha G_F^2 m_W^4}{48\pi^2} \,
\frac{ s x (1-x) }{\kappa_+ \kappa_-} \,
\left[ \eta_+^2 F(\eta_+) + \eta_-^2 F(\eta_-) \right], 
\label{sigma_gnunu}
\end{equation}
where
\begin{eqnarray}
F(\eta) &=& \frac{ N_{\nu} (g_v^2+g_a^2) + 3 (g_v+g_a)
\left(1-\frac{s(1-x)}{m_Z^2}\right) \frac{1}{\eta} \left[ 3+
\frac{2}{\eta} - 2 \left( 1+ \frac{1}{\eta} \right)^2 \log (1+\eta)
\right] }{\left(1-\frac{s(1-x)}{m_Z^2}\right)^2+
\frac{\Gamma^2_{\rm tot}}{m_Z^2}}+ \nonumber \\
&& + \frac{6}{\eta} \left[ (1+\eta) \left( 1- \frac{2}{\eta}
\log (1+\eta) \right) +1 \right], \\
\eta_\pm &=& \frac{s-\kappa_\pm}{m_W^2}, \\  
\kappa_\pm &=& \frac{s}{2} x (1 \pm y).
\end{eqnarray}
$\sqrt{s}=2E_{\rm beam}$, $E_{\rm beam}$ is the beam energy, 
$x=E_{\gamma}/E_{\rm beam}$, $x^{\rm min}=E^{\rm min}_{\gamma}/E_{\rm beam}$,
$y=\cos\theta_{\gamma}$ is the angle of the photon with respect to the beam 
direction, $g_v=-1/2+2\sin^2\theta_{W}$ and $g_a=-1/2$ are the SM vector and 
axial-vector $Ze\bar{e}$-couplings. We have assumed that the charged-current
$We\bar{\nu}$-coupling and the neutral current $Z\nu\bar{\nu}$-couplings are all equal to
their SM values. We will revisit some of these hypotheses in the next section. 
The following approximation is made when deriving Eq.~(\ref{sigma_gnunu}): the contribution
from the third diagram in Fig.~\ref{fig_diagrams}, suppressed by an extra $W$-boson
propagator, is neglected, along with the finite width of the $W$-boson (a good approximation
for ``space-like'' $W$-boson exchange).

The main SM physics backgrounds come from the processes 
$e^+e^- \to e^+e^-\gamma(n\gamma)$,
$e^+e^- \to \mu^+\mu^-\gamma(n\gamma)$,
$e^+e^- \to \tau^+\tau^-\gamma(n\gamma)$,
$e^+e^- \to \gamma\gamma\gamma(n\gamma)$ and
$e^+e^- \to l^+l^-\nu\bar{\nu}\gamma$
(see, for example, \cite{OPAL_direct}). They are characterized by a transverse tagging photon
with $E_\gamma > 1$ GeV
and additional high-energy charged particles and/or photons which are lost in
the ``blind'' regions of the detector located around the beam pipe.
The expected contributions from the process $e^+e^- \to
\nu\bar{\nu}\nu\bar{\nu}\gamma(\gamma)$ are negligible and will not be considered
henceforth. We have computed these background cross-sections using Monte Carlo
integration methods.
The number of background events can be reduced by vetoing on additional energy
deposits in the calorimeters, in particular at
low-angles. As a concrete example, we consider the TESLA detector concept
\cite{TESLA_detector}, which envisions a luminosity calorimeter (LCAL) at very small
angles ($4.6 < \theta < 27.5$ mrad). Together with the low-angle tagger (LAT)
at $27.5 < \theta < 83$ mrad, the LCAL provides an excellent angular coverage 
for the background veto. In fact, we estimate that the cross sections for 
all background sources mentioned above are reduced to a negligible level of $\sim 0.1$ fb.

A more detailed background analysis would require the inclusion of detector effects.
For example, additional contributions arise from the processes $e^+e^- \to
\nu\bar{\nu}X$ and $e^+e^- \to e^+e^-X$ with $X = \pi^0, \eta, \eta', {\rm
f}_2(1270)$ where the neutral hadron is misidentified as a photon.
However, while the contribution from the $\nu\bar{\nu}X$ cross-section is
expected to very small because of phase-space constraints,
the two-photon production of resonances in $e^+e^- \to e^+e^-X$ can be reduced
to a negligible level using the low-angle veto as discussed above. Since the
total background level is very small, additional detector effects should
therefore play a minor role.
In Table~\ref{table_stat}, 
we quote the results obtained for the signal cross section,
for different center-of-mass energies, 
including leading-log initial-state radiation and beamstrahlung using the
program \textsc{Circe} \cite{circe}.
Also given are the expected number of events
which are to be recorded after accumulating 50~fb$^{-1}$ of data in a 
Giga-Z experiment, assuming 65\% selection efficiency\footnote{This is the 
selection efficiency obtained in the OPAL analysis \cite{OPAL_direct}. It is 
slightly better than the one obtained by the L3 experiment \cite{L3_direct}} 
in the given kinematic region.
We also compute the figure of merit $1/\sqrt{S}$ for 
the different center-of-mass energies. Given that the background cross section
is well below 1~pb, we expect the number of background events to be 
negligible.
\begin{table}
\caption{Cross section for $e^+e^-\rightarrow\nu\bar{\nu}\gamma$ (signal) 
at a linear collider for three
center-of-mass energies around the $Z$-boson mass. 
Also tabulated is the expected number of
signal events, $S$, assuming that 50~fb$^{-1}$ of data are collected with an efficiency
of 65\%. See text for details.
Finally, in the last column we compute $1/\sqrt{S}$, the statistical error
which one expects to obtain when extracting the signal. 
\label{table_stat}
}
\begin{tabular}{|c|c|c|l|} \hline
$\sqrt{s}$ & $\sigma(\nu\bar{\nu}\gamma)$ & $S(\nu\bar{\nu}\gamma)$ & $1/\sqrt{S}$ \\ \hline
$m_Z=91.1875$~GeV & 53.5~pb & $1.74\times 10^6$ & 0.076\%\\ \hline
$m_Z-1$~GeV & 28.6~pb & $0.93\times 10^6$ & 0.10\%
\\ \hline
$m_Z+1$~GeV & 109~pb & $3.5\times 10^6$ & 0.053\%
\\ \hline
\end{tabular}
\end{table}
The figure of merit is the relative statistical uncertainty
for measuring the invisible $Z$-width. For a 50~fb$^{-1}$
Giga-Z experiment, one can expect statistical errors around the 0.1\%
level, about a factor 25 improvement over the statistical error quoted by L3. 

Systematic uncertainties may
dominate the very small statistical errors estimated above. In order to correctly estimate 
the systematic uncertainties, one should perform a complete detector
simulation, which is clearly beyond the intentions of this paper.
Instead, we analyze the systematic errors that were
computed by the LEP experiments for the same measurement, and extrapolate
them for a TESLA-like Giga-Z experiment.
We concentrate mostly on the L3 1998 systematic error computations,
obtained from the analysis of 100~pb$^{-1}$ of data \cite{L3_direct}. 
These are presented in Table~\ref{table_sys}. For illustrative purposes, we also 
quote the systematic errors computed in earlier analyses by ALEPH (which
analyzed 19~pb$^{-1}$ of data \cite{ALEPH_direct}) and OPAL (based on
40.5~pb$^{-1}$ of data \cite{OPAL_direct}). We make
use of these results to verify that our estimates are reasonable. 

\begin{table}
\caption{Systematic uncertainties for measuring the invisible $Z$-width, in 
percent and (inside the square brackets) expressed as 
$\delta\Gamma_{\rm inv}$. The source of the systematic uncertainty is listed
in the first column (see text for details) while the second through fourth
columns contain the estimates obtained by ALEPH \cite{ALEPH_direct} in 1993 
(19~pb$^{-1}$ of data),
OPAL \cite{OPAL_direct} 1995 (40.5~pb$^{-1}$ of data) and L3 \cite{L3_direct} 
1998 (100~pb$^{-1}$ of data). Our projection for TESLA running
in the Giga-Z mode (50~fb$^{-1}$) is presented in the last column. N/C indicates that this
source of systematic error was not considered or not quoted in the specific published
result.
\label{table_sys}
}
\begin{tabular}{|c|c|c|c||c|} \hline
Source of Systematic Error & ALEPH 93 & OPAL 95 & L3 98 & TESLA (estimate)
\\ \hline
event generator for $\nu\bar{\nu}\gamma$ & 1\% [5~MeV] & 1.2\% [6~MeV] &
0.7\% [3.5~MeV] & 0.1\% [0.5~MeV] \\ \hline
event generator for $e^+e^-\gamma$ & 1\% [5~MeV] & in bkgd. subtr. &
0.7\% [3.5~MeV] & 0.1\% [0.5~MeV] \\ \hline
energy calibration & 1.5\% [7.5~MeV] & 1.7\% [9~MeV] &
0.8\% [4~MeV] & 0.03\% [0.15~MeV] 
\\ \hline
luminosity & 0.6\% [3 MeV] & 0.6\% [3 MeV] & 0.37\% [1.8~MeV] &
0.06\% [0.3~MeV] \\ \hline
fit procedure & N/C & 0.9\% [5 MeV] &
0.5\% [2.5~MeV] & 0.1\% [0.5~MeV] \\ \hline
selection efficiency and & 3.9\% [18~MeV] & 1.7\% [9~MeV] &
0.8\% [4~MeV] & $<$0.08\% [0.4~MeV] 
\\ 
veto efficiency & 1.8\% [9~MeV] & 0.5\% [2.5~MeV] & &
\\ \hline
trigger efficiency & 0.2\% [1~MeV] & 0.1\% [0.5~MeV] &
1\% [4.8~MeV] & 
0.01\% or 0.04\% [(0.05 or 0.21)~MeV]  \\ \hline
background subtraction & N/C & 1.6\% [8~MeV] &
1.7\% [8.4~MeV] & negligible  
\\ \hline
cosmic ray background & N/C & in bkgd. subtr. & 0.25\% [1.7~MeV] & negligible 
\\ \hline
random vetoing (occupancy) & 0.5\% [2.5~MeV] & 0.5\% [2.5~MeV] & N/C & negligible 
\\ \hline \hline
total error (added in quadrature) & 6.8\% [34~MeV] & 3.3\% [17~MeV] &
2.5\% [12.3~MeV] & 0.20\% 
[1~MeV]  
\\ \hline
\end{tabular}
\end{table}

Most of the systematic
uncertainties go down simply because the number of events goes up. 
The same trend is observed when one compares the L3 result with 
the older results from the other LEP experiments. This can be appreciated, for example, 
by looking at columns 2 and 3 in Table~\ref{table_sys}. 
\begin{itemize}

\item{
By ``event generators'' we refer to the numerical accuracy of the computation of 
the signal and the background given a set of kinematical constraints. 
We expect that these theoretical 
calculations will improve by a factor of roughly 10 by the time a Giga-Z experiment is ready
to take data.}

\item{
The ``energy calibration'' of the experiment is crucial for measuring the photon energy 
and hence the lower bound $E_{\gamma}^{\rm min}$ defined above. The 
improvement suggested in the table can be achieved by calibrating the photon energy via
a comparison of other processes that yield a photon, such as 
$e^+e^-\rightarrow\ell^+\ell^-
\gamma$, $e^+e^-\rightarrow X\pi^0\rightarrow X\gamma\gamma$, etc. The calibration error
will decrease with an increase in the statistical sample 
({\it i.e.,}\/ proportional to $1/\sqrt{N}$). Since we
expect 500 times more events at the Giga-Z experiment compared to LEP, the error should
improve by a factor $\sqrt{500}\simeq22$.}

\item{
The luminosity is obtained through the 
measurement of  Bhabha scattering. Given that the cross section for Bhabha
scattering around the $Z$-boson mass is $\sigma_{\rm Bhabha}\simeq 50$~nb, 
one expects a tiny statistical 
error on the luminosity measurement of $\delta_{\rm lum}^{\rm stat}\simeq\pm0.0025\%$.
The systematic 
error constrained by luminosity monitoring has been studied for the TESLA proposal by 
\cite{TESLA_study}, and is given by $\delta_{\rm lum}^{\rm syst}=\pm0.03\%\pm0.05\%$, where 
the first number is related to experimental systematic effects, while the second one to 
theoretical effects, including beamstrahlung, etc. Combining the three errors in quadrature, 
one obtains $\delta_{\rm lum}\simeq\pm0.06\%$.}

\item{
By ``fit procedure'' we mean the error which comes from the uncertainties of other input 
physics parameters needed in order to extract the invisible $Z$-width and 
estimate the background level. These include $m_Z$, $\Gamma_{\rm tot}$, and $\Gamma_{ee}$. 
Using the latest combined results from LEP, we expect a factor of 5 improvement with 
respect to the L3 analysis, while the Giga-Z data itself should provide an extra factor of 
2 improvement on $\Gamma_{ee}$.}

\item{
The ``selection and veto efficiencies'' are estimated via a comparison of data and 
Monte Carlo. The uncertainty is partially controlled by the size of the data 
sample, so again we can expect a factor $\sqrt{500}\simeq22$ 
improvement of both of these systematic uncertainties. The other contribution to the 
uncertainty comes from the quality of the Monte Carlo simulations, which we assume will 
improve by roughly a factor of 10.}

\item{
The ``trigger efficiency'' can be studied via control samples with independent
triggers ({\it e.g.}\/ hadronic events, $\ell^+\ell^-\gamma$, etc), indicating
that the systematic error is also related to the overall data sample.
We note that the L3  estimate presented in the fourth column of
Table~\ref{table_sys} is much larger than the ALEPH or OPAL numbers
presented in the second and third column. We therefore quote two estimates  for
the trigger efficiency, one based on the L3 estimate and one on the ALEPH
estimate.}

\item{After applying selection and kinematic cuts, some background contribution
remains, and it needs to be subtracted. The precise value of the remaining
background crucially depends on the performance of the detector in rejecting
charged particles at small angles. The understanding of the detector
systematics in this region is afflicted with some systematic uncertainty. 
We conservatively attribute an error of 20\% to the computation of
the background contamination. Note, however, that since background levels can be
reduced to negligible levels in the presence of a luminosity calorimeter, the impact of
this uncertainty is negligible.}

\item{
The much higher luminosity of a Giga-Z machine should render cosmic rays irrelevant.  
The impact of detector and beam-related noise can be estimated
with special ``zero-bias'' triggers, with negligible uncertainty.}

\end{itemize} 

In summary, we estimate the combined systematic error to be around 
$(\delta\Gamma_{\rm inv})^{\rm sys}\simeq \pm1$~MeV. Due to the disparity among
different LEP measurements, we can, in principle, quote a ``best'' and ``worst'' 
case scenario. In the best case, the trigger efficiency is $\pm0.01$\% uncertain, 
while in the worst case, 
the trigger efficiency is measured with a $\pm0.05$\% error. 
In practice, however, the ``best'' and
``worst'' cases yield the same total systematic error. 
Note that all uncertainties have been added in quadrature. 

Our estimate of the total systematic error 
is already a factor of two larger than the statistical error estimated earlier, 
$(\delta\Gamma_{\rm inv})^{\rm stat}\simeq \pm0.5$~MeV, so the question of
whether the accumulation of many more events would lead to a significant 
improvement of the measurement requires a more detailed analysis. The overall error, 
$\delta\Gamma_{\rm inv}\simeq \pm1.3$~MeV, is slightly smaller than the one obtained at 
LEP via the indirect method, Eq.~(\ref{LEP_res}),
and is comparable to the estimated indirect result that might be obtained by the
Giga-Z experiment itself ($\delta(\Gamma_{\rm inv})=\pm(0.5~{\rm to}~1.1)$~MeV). More
importantly, the direct measurement at the Giga-Z experiment is expected to be a factor
15 times more precise than the current direct measurement obtained by the four LEP 
collaborations and a factor of roughly 6 times more precise than the ultimate 
precision that can be reach by analyzing the entire LEP data set. 
Finally, a  
combined result (if one could be properly defined) would have an error bar that is similar 
to the current theoretical uncertainty in calculating the partial width for 
$Z\rightarrow\nu\bar{\nu}$ in the SM.

\setcounter{footnote}{0}
\setcounter{equation}{0}
\section{$Z\nu\bar{\nu}$-couplings away from the $Z$-pole}
\label{section_away}

At any center-of-mass energy, the differential cross section for $e^+e^-\rightarrow
\gamma\nu\bar{\nu}$, in the SM, assuming generic $Z\nu\bar{\nu}$-couplings and
neglecting neutrino-mass effects
is given by
\begin{equation}
\frac{{\rm d}\sigma_{\gamma\nu\bar{\nu}}}{{\rm d}x}=\left(\sum_{\alpha=e,\mu,\tau}
\left[(g_L^{\nu_{\alpha}})^2+(g_R^{\nu_{\alpha}})^2\right]ZZ(s,x)\right)+
\left(g_L^{\nu_e}\right)WZ(s,x)+
WW(s,x).
\label{sigma_general}
\end{equation}
The leading order Feynman diagrams are shown in
Fig.~\ref{fig_diagrams}. Here $x$ and $s$ are defined as in Eq.~(\ref{sigma_gnunu}),
while $ZZ,~WW$, and $ZW$ are functions of $s,x$ (plus several standard model parameters, 
including $m_Z^2,~\Gamma_{\rm tot}$, and the $Ze\bar{e}$-couplings). 
The first term corresponds to the square of the $s$-channel 
$Z$-boson exchange amplitude and the third term
to the square of the $t$-channel $W$-boson exchange amplitude, while the second term
arises from the interference between these two contributions. We have 
made explicit the dependency on the $Z\nu_{\alpha}\bar{\nu}_{\alpha}$ 
left-handed and right-handed couplings ($\alpha=e,\mu,\tau$). In the SM model, 
$g_L^{\nu_e}=g_L^{\nu_{\mu}}=g_L^{\nu_{\tau}}=1/2$, while 
$g_R^{\nu_e}=g_R^{\nu_{\mu}}=g_R^{\nu_{\tau}}=0$.
We assume throughout that the charged-current $We\bar{\nu}_e$-coupling agrees with
its SM prediction. Experimentally, the charged-current neutrino-electron coupling
is well constrained to be purely left-handed (at the few percent level), 
and its value is accurately determined. Needless to say, the 
$W\ell\bar{\nu}$-couplings are much better constrained (directly) than the 
$Z\nu\bar{\nu}$-couplings.
The most stringent constraints on the nature and value of the $W\ell\bar{\nu}$-couplings 
are provided by studying weak decays of neutrons, nuclei, muons, and charged pions. 
We refer readers to, for example, \cite{PDG,v-a} for details.

At an $e^+e^-$-collider it is impossible to distinguish $\nu_{\tau}\bar{\nu}_{\tau}$
from $\nu_{\mu}\bar{\nu}_{\mu}$ final states, which allows one to rewrite the coefficient
of $ZZ$ in Eq.~(\ref{sigma_general}) as
\begin{equation}
\sum_{\alpha=e,\mu,\tau}
\left[(g_L^{\nu_{\alpha}})^2+(g_R^{\nu_{\alpha}})^2\right]\equiv N_{\nu}
\left[(g_L^{\nu_{e}})^2+(g_R^{\nu_{e}})^2\right],
\label{define_gr_gl_Nnu}
\end{equation}
where $N_{\nu}$ is the effective neutrino number. 
This definition of $N_{\nu}$ only agrees
with the one in Eq.~(\ref{number_of_nus}) if 
the charged-lepton couplings to the $Z$-boson are fixed to their SM values. 
As a matter of fact, the right-handed $Z\nu\bar{\nu}$-coupling $g_R$ 
can be more generally interpreted as coupling of the $Z$-boson to other exotic, 
invisible final states. We will return to this point in Sec.~\ref{section_theory}.   

\begin{figure}[t]
{\centering
\epsfig{figure=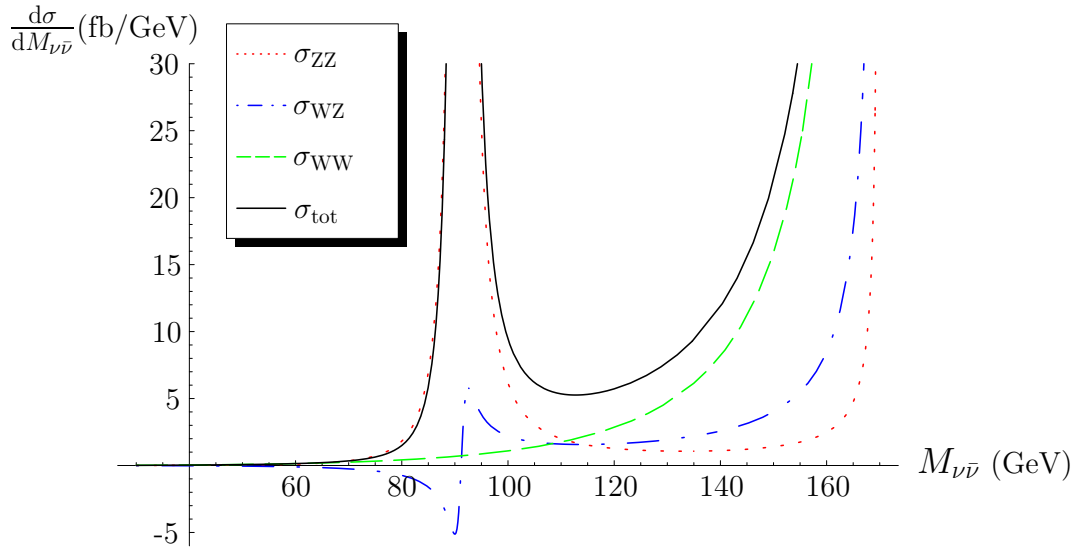,width=13cm}
}
\caption{The differential cross-section for $e^+e^-\rightarrow\gamma\nu\bar{\nu}$ as a 
function of the invariant mass of the $\nu\bar{\nu}$-system, $M_{\nu\bar{\nu}}$, 
for $\sqrt{s}=170$~GeV, assuming SM values for the number of neutrino species and the 
neutrino neutral-current couplings. We also show the different contributions to the
differential cross section ($ZZ,WZ$, and $WW$ -- see text for details). The sharp increase
of the differential cross section as $M_{\nu\bar{\nu}}$ approaches $\sqrt{s}$ is due to 
an infrared singularity at vanishing photon energy.
\label{fig_comparison}
}
\end{figure}

In order to analyze the kinematics of $e^+e^-\rightarrow\gamma\nu\bar{\nu}$, it 
proves useful to utilize the ``missing mass,'' defined to be the 
mass of the system recoiling against the photon:
$M_{\nu\bar{\nu}}\equiv\sqrt{s(1-x)}$.
If there are no additional photons, this coincides with the 
$\nu\bar{\nu}$~invariant mass.
For missing mass close to the $Z$-boson
mass, the cross section for $e^+e^-\rightarrow\gamma\nu\bar{\nu}$ is dominated
by the $ZZ$-term, and one can only, in practice, measure 
$N_{\nu}(g_L^2+g_R^2)$.\footnote{Henceforth, we replace $g_{L,R}^{\nu_e}$ with $g_{L,R}$.} 
On the other hand, for a range of values of
the missing mass above the $Z$-boson mass
(or the photon energy, $E_{\gamma}=xE_{\rm beam}$, below the $Z$-boson mass), 
the $ZZ$, $WZ$ and $WW$
contributions are comparable and one is, in principle, sensitive to {\sl both} $g_L$
and $N_{\nu}(g_L^2+g_R^2)$. For very high values of $M_{\nu\bar{\nu}}$, however, the $WW$-term 
dominates, and one loses sensitivity to both $N_{\nu}(g_L^2+g_R^2)$ and $g_L$.
Fig.~\ref{fig_comparison} depicts 
${\rm d}\sigma_{\gamma\nu\bar{\nu}}/{\rm d}M_{\nu\bar{\nu}}$ as a function of $M_{\nu\bar{\nu}}$, 
for $\sqrt{s}=170$~GeV, $N_{\nu}=3$ and the SM values for the neutral-current 
neutrino couplings. Fig.~\ref{fig_comparison} also displays the different contributions to 
the differential cross section. As one can easily note, for $M_{\nu\bar{\nu}}$ around the $Z$-boson
mass the differential cross section is completely dominated by the $ZZ$-term, while
at the largest values of $M_{\nu\bar{\nu}}$ the $WW$-piece dominates. The interference $WZ$-term, which
changes sign at the $Z$-boson mass, becomes comparable to the other two contributions 
at $M_{\nu\bar{\nu}}\sim 100$~GeV. 
For a fixed value of $N_{\nu}$, one can therefore measure $g_L$ directly by
measuring the cross section for $e^+e^-\rightarrow\gamma\nu\bar{\nu}$ above
$M_{\nu\bar{\nu}} \sim 100$~GeV.
Before pursuing this further, however, we will first
review what is currently known about the values of $g_L$ and $g_R$.

\subsection{Current Knowledge of the $g^{\nu}_L$ and $g^{\nu}_R$ Couplings to the $Z$-boson}

\begin{figure}[t]
{\centering
\epsfig{figure=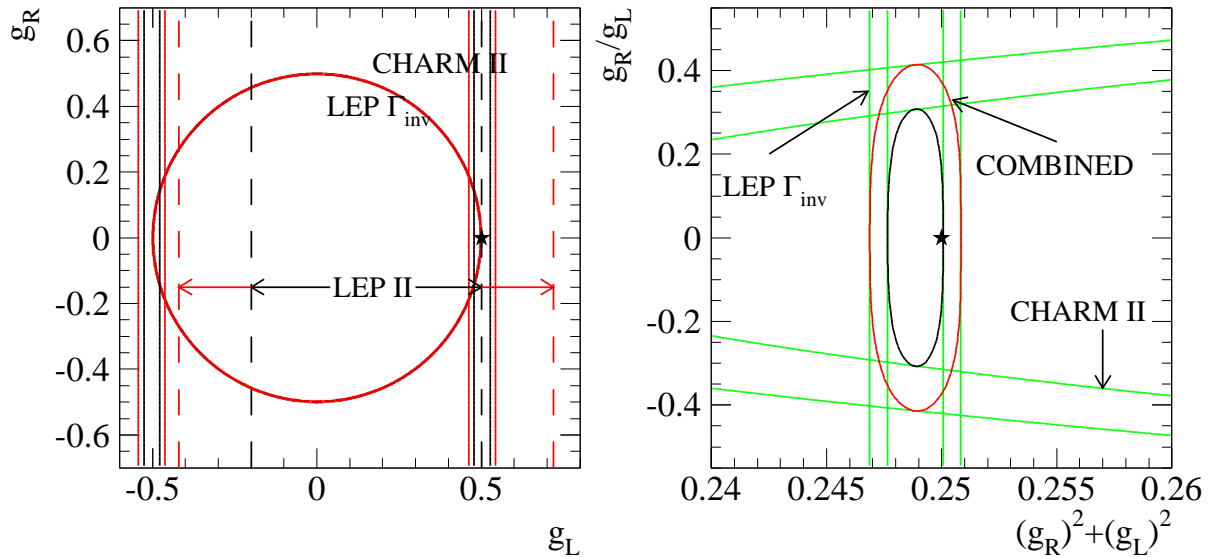,width=0.9\textwidth}
}
\caption{(left)-Current constraint (one and two sigma confidence level contours)
on $g_L$ and $g_R$ from the LEP (indirect) measurement of
the invisible $Z$-width and the CHARM~II experiment, on the $g_L\times g_R$-plane. The 
SM expectation is indicated by a star. We have assumed $N_{\nu}=3$ for the LEP result, 
and $g_L=g_L^{\nu_{\mu}}$ for the CHARM~II result in order to have both experiments
constrain the same physical parameters. We also include the current constraint on $g_L$
than can be obtained from published LEP~II data, also at the one and two sigma
confidence level. See text for details. (right)-The one and two
sigma allowed regions which are selected by combining the LEP and CHARM~II results. The individual
LEP and CHARM~II constraints (at one and two sigma confidence levels) are also depicted. 
Note that one is unable to distinguish $g_L>0$ from $g_L<0$.
\label{fig_LEPcharm2}
}
\end{figure}

The currently most precise value of $N_{\nu}(g_L^2+g_R^2)$ can be extracted from 
the indirect measurement of the invisible $Z$-width, Eq.~(\ref{LEP_res}). 
For $N_{\nu}=3$, the region of the $g_L\times g_R$-plane allowed by Eq.~(\ref{LEP_res}) 
is characterized  by a ring. Fig.~\ref{fig_LEPcharm2}(left) shows the current LEP constraint
at one and two sigma confidence levels (the two contours are indistinguishable
in the figure). 

More information is provided by $\nu-e$-elastic scattering experiments. The CHARM~II 
experiment at CERN collected a large sample of $\nu_{\mu}e\rightarrow \nu_{\mu}e$ and 
$\bar{\nu}_{\mu}e\rightarrow \bar{\nu}_{\mu}e$ events \cite{charm2}. 
By using information on the $Ze\bar{e}$
vector and axial vector couplings \cite{charm2theory} 
measured very accurately at LEP and SLC (see, for example, \cite{LEP_Z,combined,combined_2} 
and references therein), 
CHARM~II is capable of measuring $|g_L^{\nu_{\mu}}|$ rather well:
\begin{equation}
|g_L^{\nu_{\mu}}| = 0.502\pm0.017 \qquad ({\rm CHARM~II}),
\label{gL_CHARMII}
\end{equation}   
where we quote the updated number presented by the PDG \cite{PDG}. Furthermore, CHARM~II 
can also measure $|g_L^{\nu_e}|({\rm CHARM~II})=0.528\pm0.085$ via a small 
``contamination'' of
$\nu_{e}(\bar{\nu}_{e})e\rightarrow\bar{\nu}_e(\bar{\nu}_e)e$ 
events which are present in its data set.
This result agrees with the one obtained in a   
$\nu_e-e$-elastic scattering experiment at the Los Alamos Meson Physics Facility (LAMPF): 
$|g_L^{\nu_e}|({\rm LAMPF})=0.46\pm0.14$ \cite{LAMPF}. 
In order to claim that these neutrino--electron scattering experiments are indeed sensitive only
to the left-handed $Z\nu\bar{\nu}$-couplings, we are assuming that the charged-current
interactions responsible for producing the neutrino beam are purely 
left-handed and neutrino-mass effects can be neglected. The region of the 
$g_L\times g_R$-plane allowed by Eq.~(\ref{gL_CHARMII}) is depicted as vertical bars (at one
and two sigma confidence level) in Fig.~\ref{fig_LEPcharm2}(left). 
The SM value for $(g_L,g_R)_{\rm SM}$ is represented by a star. 
 
In order to combine the LEP invisible $Z$-width constraint with the CHARM II bound, 
it is useful to display the result on the $(g_L^2+g^2_R)\times g_R/g_L$-plane. 
In this case,  the region depicted in Fig.~\ref{fig_LEPcharm2}(right)
is selected at one and two sigma confidence level. Here, the region allowed by the
invisible $Z$-width measurement at LEP is characterized by vertical bars, while the CHARM II
bound is characterized by a ``parabolic'' region. 
It is important to emphasize that in order for this joint analysis to make sense, we are 
assuming that the $Z\nu\bar{\nu}$-couplings are universal (the same for all three 
neutrino flavors), that there are three neutrino species (coupling to the $Z$-boson), 
and that there are no extra contributions to invisible $Z$-boson decays or 
electron--neutrino scattering. The result obtained is rather
good: $|g_L|$ has been measured with relatively good precision 
($0.45\lesssim |g_L| \lesssim 0.5$). On the other hand, $|g_R|$ is only mildly bounded 
from above ($|g_R|\lesssim 0.2$), and we have no information concerning the sign of $g_L$.  

The NuTeV experiment also provides a measurement of the muon-neutrino coupling to the
$Z$-boson. Assuming the value of $\sin^2\theta_W$ obtained at other experiments, 
SM values for the $Zq\bar{q}$-couplings, and
fixing the $W\mu\bar{\nu}_{\mu}$-coupling to its SM value,
one can interpret the NuTeV result as a measurement of $|g_L^{\nu_{\mu}}|$. From
\cite{NuTeV2}, 
\begin{equation}
|g_L^{\nu_{\mu}}|=0.4971\pm0.0011  \qquad ({\rm NuTeV}).
\label{gL_NuTeV}
\end{equation} 
This result is 15 times more precise than the CHARM~II result (Eq.~(\ref{gL_CHARMII})), 
and 1.4 times more precise than the LEP result (Eq.~(\ref{LEP_res})). Furthermore, while its central
value is roughly $3\sigma$ away from the SM prediction, Eq.~(\ref{gL_NuTeV}) is perfectly (within
one sigma) consistent with Eq.~(\ref{LEP_res}), which also differs from the
SM prediction by 1.5$\sigma$, 
and (Eq.~(\ref{gL_CHARMII})), which is a lot less precise. Nonetheless, 
we choose not include it in our studies $g_L$ and $g_R$,
for a few reasons. Many assumptions have to be made before one can interpret the 
NuTeV result as a measurement of the $Z\nu\bar{\nu}$-coupling, including
the assumption that the $Z$-boson coupling to quarks is as prescribed by the SM. 
In the case of the CHARM~II result, in contrast, we only had to input the values of 
$g_v$ and $g_a$ which were directly measured at LEP. 
More importantly, there is a significant amount of discussion in the literature 
concerning whether nuclear and/or hadronic effects might further modify 
the NuTeV result (see, for example, \cite{NuTeV_review}) and it is still premature to 
compare Eq.~(\ref{gL_NuTeV}) with 
the other measurements of the $Z\nu\bar{\nu}$-couplings discussed earlier.   

We return now to Eq.~(\ref{sigma_general}) and investigate the impact of the LEP~II data.  
These were collected at different center-of-mass energies above the $Z$-boson mass,
but not all of have been used to  measure the cross sections for
$e^+e^-\rightarrow\gamma\nu\bar{\nu}$.  A useful summary is given in Ref.~\cite{Hirsch}.
In order to extract $g_L$, we compute the total cross section at each $\sqrt{s}$
imposing the various fiducial and kinematic cuts of each measurement.  The coefficient 
for the $ZZ$-term in Eq.~(\ref{sigma_general}) is constrained to the value obtained 
from $\Gamma_{\rm inv}^{LEP}$, Eq.~(\ref{LEP_res}). We construct an overall $\chi^2$ 
function, taking all systematic uncertainties to be wholly correlated.  Since the 
measurement errors are dominated by the statistical uncertainties, the correlations 
are not very important numerically.  Minimization of $\chi^2$ gives
\begin{equation}
  g_L = 0.16 \pm 0.23  \qquad ({\rm LEP~II}),
\label{gL_LEPII}
\end{equation} 
and $\chi^2 = 9.6$ for 23 degrees of freedom.  The allowed range of $g_L$ is indicated
in Fig.~\ref{fig_LEPcharm2}(left) at the one and two sigma confidence levels\footnote{
In order to facilitate comparison with the other measurements, the error on $g_L$ has
been rescaled to correspond to two free parameters.}  These data do eliminate the
region $g_L \approx -\frac{1}{2}$, otherwise allowed by the CHARM~II and LEP~I data,
at a little more than the two sigma confidence level.

The LEP~II data could provide a much stronger constraint.  First of all, only ALEPH
has published measurements from its entire data sample.  If the other collaborations
completed the analysis of all data with $\sqrt{s} > 160$~GeV using the fiducial
and kinematic cuts of the published measurements, we estimate that the one-parameter error on 
$g_L$ would decrease to $\delta(g_L) = 0.15$.  More importantly, the published
LEP~II analyses are not optimized for measuring $g_L$.  As is apparent from
Fig.~\ref{fig_comparison}, events with a missing mass close to the $Z$-pole mass
will dominate the total cross section and dilute the impact of the interference
term, $WZ$.  We estimate that the imposition of  
a lower limit on the missing mass should improve the sensitivity to $g_L$
by about a factor of three.  The optimal value for this cut is around $95$--$100$~GeV.
If all LEP~II data were analyzed with this cut imposed, the total error on~$g_L$
should decrease to $\delta(g_L) = 0.05$ -- better than a factor four improvement.  
Of course, the central value in Eq.~(\ref{gL_LEPII}) is also likely to change.  
Only an analysis of the actual data taken by the LEP collaborations will reveal it.

\subsection{Measuring $g^{\nu}_L$ and $g^{\nu}_R$ in a Linear Collider}

\begin{figure}[t]
{\centering
\epsfig{figure=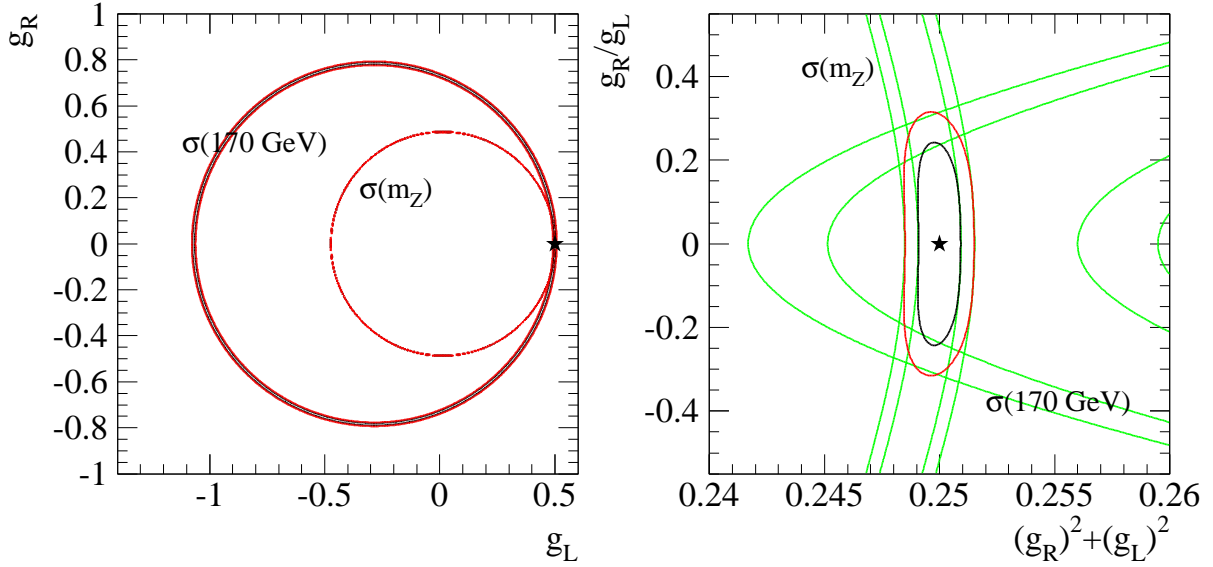,width=0.9\textwidth}
}
\caption{(left)-Projected constraint (one and two sigma confidence level contours)
on $g_L$ and $g_R$ from  $e^+e^-\rightarrow\gamma\nu\bar{\nu}$ at a linear collider assuming
50~fb$^{-1}$ of data are collected at the $Z$-boson mass ($\sigma(m_Z)$) and at 
$\sqrt{s}=170$~GeV ($\sigma(170~{\rm GeV})$), on the $g_L\times g_R$-plane. 
For the result obtained at $\sqrt{s}=170$~GeV, we impose
a constraint on the invariant mass of the $\nu\bar{\nu}$-system ($M_{\nu\bar{\nu}}>100$~GeV) in order 
to remove the radiative return to the $Z$-boson mass. 
(right)-The one and two sigma allowed regions which are selected by combining the 
results obtained at the two different center of mass energies
in the $(g_R^2 + g_L^2) \times g_R/g_L$-plane. 
The individual constraints obtained at the two distinct center-of-mass energies are 
also depicted.
In both figures,
the SM expectation is indicated by a star and we have assumed $N_{\nu}=3$.
\label{fig_LC}
}
\end{figure}

We now discuss how a linear collider can improve on the existing 
results discussed above. In order to do this we compute the 
$e^+e^-\rightarrow \gamma\nu\bar{\nu}$ cross section at a
linear collider at $\sqrt{s}=m_Z$ and $\sqrt{s}=170$~GeV. 
The latter collider center-of-mass energy can also be used, for example, for precisely 
measuring the $W$-boson mass \cite{LC_WW}.\footnote{For our purposes, the choice of the 
``high'' center-of-mass energy is not crucial. Any value in the range $[150-200]$~GeV 
will yield similar results.}
Assuming that the central value for $\Gamma_{\rm inv}$ obtained at the Giga-Z machine agrees 
with the SM prediction, we can translate an expected $\pm 0.25$\% uncertainty, as estimated in 
Sec.~\ref{section_Zpole}, into an allowed region of the $g_L\times g_R$-plane. 
This region is characterized by a ring approximately centered around the origin, and is 
depicted in Fig.~\ref{fig_LC}(left). The shape and width
of the curve are almost identical to the one depicted in Fig.~\ref{fig_LEPcharm2}(left) 
for the LEP indirect measurement of the invisible $Z$-width. Upon closer inspection, one should
be able to see that the center of the ring is slightly shifted to the right. This small effect 
is due to the nonzero contribution of the $W$-boson exchange diagram. 
Furthermore, the precision 
obtained from the direct measurement of the invisible $Z$-width at Giga-Z is only slightly better 
than the current indirect LEP result, and slightly worse than the future indirect 
result that might be obtained by Giga-Z (see Sec.~\ref{section_Zpole}). However, it is interesting
to discuss with what precision the neutrino neutral-current couplings can be measured at a linear
collider when one compares the same observable (namely, the cross section for 
$e^+e^-\rightarrow\gamma+$~invisible) measured at different center-of-mass energies. By doing this,
we reduce the number of assumptions that go into extracting $g_L$ and $g_R$, and potentially
minimize experimental ``biases'' that may affect different observables in different ways.  

We compute the $e^+e^-\rightarrow \gamma\nu\bar{\nu}$ cross section using 
Eq.~(\ref{sigma_gnunu})~\cite{sigma_gnn}. As before, we require $E_{\gamma}>1$~GeV,
and $\theta_{\gamma}>20^{\circ}$. In order to enhance the sensitivity to the 
$WZ$ interference term (see Eq.~(\ref{sigma_general})) 
we also require  $M_{\nu\bar{\nu}}>100$~GeV .
Assuming SM values for the $Z\nu\bar{\nu}$-couplings and $N_{\nu}=3$, we obtain
\begin{equation}
\sigma_{\gamma\nu\bar{\nu}}(\sqrt{s}=170~{\rm GeV},M_{\nu\bar{\nu}}>100~{\rm GeV})=2.97~{\rm pb}.
\end{equation}
Assuming 50~fb$^{-1}$ of linear collider data and an efficiency of 80\% \cite{OPAL_gnn,ALEPH_gnn}, 
we expect around $120\,000$
$e^+e^-\rightarrow\gamma\nu\bar{\nu}$ events with $M_{\nu\bar{\nu}}>100$~GeV. 

The dominant sources of SM physics background have been listed in Sec.~\ref{section_Zpole}B 
(see also \cite{OPAL_gnn}) and can be dramatically reduced by vetoing on additional
high-energy particles, as discussed earlier. Including the LCAL
of the TESLA detector design \cite{TESLA_detector} 
which helps veto hard particles at very low angles, 
the total cross section for these backgrounds can be reduced to less than 1~fb.

Other
important background sources can arise from detector-related effects. As
discussed in \cite{OPAL_gnn}, the dominant background contributions are related to
the processes $e^+e^- \to \nu\bar{\nu}\ell^+\ell^-$ and $e^+e^- \to e^+e^-\ell^+\ell^-$,
which can mimic events where a photon converts into a lepton pair in the
material of the detector. We adopt the values quoted in \cite{OPAL_gnn} of $(0.010 \pm
0.001)$~pb and $(0.007 \pm 0.002)$~pb, respectively, for the cross sections for the above
background processes. In order to accommodate any changes in detector design compared to 
the OPAL detector \cite{OPAL_gnn}, we conservatively allow a factor of two uncertainty 
on these cross sections, resulting in a total background cross section of about 0.03 pb.

Assuming an integrated luminosity of 50~fb$^{-1}$, the estimated number of
background events is $N_{\rm bkg}=1500$. Note that this number, which is 
very conservative, also includes events with $M_{\nu\bar{\nu}}<100$~GeV, which have been 
removed when we estimate the number of signal events. Given the values for the
cross sections computed above, we estimate (conservatively) that the statistical uncertainty 
which can be achieved
after accumulating 50~fb$^{-1}$ of $e^+e^-\rightarrow\gamma+$~invisible at $\sqrt{s}=170$~GeV
is $\sqrt{S+B}/S=0.3$\%.

Given the very small statistical errors estimated above, we must try to evaluate 
the size of possible systematic uncertainties. Following the strategy outlined in 
Sec.~\ref{section_Zpole}B, we analyze the systematic errors that were computed by
the different LEP collaborations for the same observable, and extrapolate them for a 
TESLA-like linear collider. This time, we concentrate on the analyses of 177~pb$^{-1}$ of data
collected by OPAL \cite{OPAL_gnn} and 628~pb$^{-1}$ of data collected by ALEPH \cite{ALEPH_gnn}.
Their estimates for different systematic uncertainties are presented in Table~\ref{table_gnn},
together with our extrapolation for a linear collider experiment accumulating
50~fb$^{-1}$ of data.
There are similar analyses by L3 \cite{L3_gnn} and DELPHI \cite{DELPHI_gnn}, but their
discussions of the systematic errors are not as detailed as the previous two.

\begin{table}
\caption{Systematic uncertainties for measuring the cross-section for
$e^+e^-\rightarrow\gamma\nu\bar{\nu}$ at center-of-mass energies above $\sim160$~GeV, in 
percentage. The source of the systematic uncertainty is listed
in the first column (see text for details) while the second and third
columns contain the estimates obtained by OPAL \cite{OPAL_gnn} (177~pb$^{-1}$ of data) and
ALEPH \cite{ALEPH_gnn} (628~pb$^{-1}$ of data). Our projection for TESLA running
at $\sqrt{s}=170$~GeV (and collecting 50~fb$^{-1}$ of data) 
is presented in the last column. N/C indicates that this
source of systematic error was not considered or not explicitly quoted.
\label{table_gnn}
}
\begin{tabular}{|c|c|c||c|} \hline
Source of Systematic Error & OPAL (177~pb$^{-1}$) & ALEPH (628~pb$^{-1}$) & TESLA [estimate] 
(50~fb$^{-1}$)
\\ \hline
event generator -- theoretical & 0.5\% & 1.5\% &
0.2\% \\ \hline
event generator -- statistical & 0.2\% & 0.5\% &
$<$0.1\% \\ \hline
energy calibration & 0.4\% & N/C &
0.025\% \\ \hline
luminosity & 0.2\% & 0.5\% & 0.06\% \\ \hline
uncertainty from $W/Z$-boson mass & N/C & N/C &
0.25\% \\ \hline
selection efficiency & 1.5\% & 0.6\% & 0.07\% \\ \hline
angular acceptance & 0.2\% & N/C & 0.01\% \\ \hline 
modeling early $\gamma$ conversion & 0.7\% & 0.3\% & 0.2\% \\ 
in material near beam-pipe & & & \\ \hline
tracking & 0.5\% & N/C & 0.1\% \\ \hline
\hline
total error (added in quadrature) & 2.1\% & 1.8\% & 0.4\% \\ \hline
\end{tabular}
\end{table}

We now briefly discuss the origin of the different systematic uncertainties, and how our
estimates were obtained. Some of
the systematic errors are related to the size of the data sample. Whenever this is the case, we
expect a factor $\sqrt{50000/177}\simeq 17$ ($\sqrt{50000/628}\simeq 9$) improvement
with respect to the OPAL (ALEPH) estimate.
\begin{itemize}

\item{As before, we expect the theoretical uncertainty 
(``event generator -- theoretical'') to improve by a factor of~10. The
statistical errors associated with these computations are only limited by the computer power
which is available to perform such numerical calculations, and we expect them to be negligible
by the time this experiment takes data.}

\item{As estimated before, the uncertainty related to energy calibration should be controlled
by statistics. The same applies for the selection efficiency and the angular acceptance. }

\item{We assume the luminosity uncertainty to be the same as the one estimated in 
subsection \ref{section_Zpole}B (see \cite{TESLA_study}). It should be emphasized, however,
that the studies performed in \cite{TESLA_study} concentrated on center-of-mass energies
around the $Z$-boson mass. We are assuming that similar numbers will apply for higher 
center-of-mass energies.}

\item{There are intrinsic uncertainties in computing the signal and background from the finite
accuracy of the input electroweak parameters (of special importance are the values of the
$W$ and $Z$-boson masses). We assume $\delta(m_Z)=\pm2$~MeV
(from LEP) and $\delta(m_W)=\pm 15$~MeV (from the linear collider itself 
and the LHC \cite{W_mass_future}).\footnote{Some studies suggest that $\delta(m_W)=\pm 6$~MeV
could be obtained by scanning around the $W^+W^-$-production threshold region \cite{LC_WW}.} 
This source of systematic uncertainty was not considered in the LEP analyses 
\cite{OPAL_gnn,ALEPH_gnn,L3_gnn,DELPHI_gnn}. By taking the combined LEP result for the 
$W$-boson mass {\sl at the time of the analysis},\/ 
($\delta(m_W)=\pm0.056$~GeV \cite{combined_old} and 
$\delta(m_W)=\pm0.042$~GeV \cite{combined_2} 
we obtain a systematic error of 0.9\% for the OPAL analysis and 0.7\% for the 
ALEPH analysis, respectively. 

\item{Photons can convert to charged particles in the material which surrounds 
the beam-pipe. This conversion rate is estimated by modeling the material close 
to the beam-pipe, and depends on the details of the detector layout. A substantial 
reduction of this error was obtained between the OPAL \cite{OPAL_gnn} and 
ALEPH \cite{ALEPH_gnn} analyses, and we assume that at  least an extra 50\% 
improvement can be obtained.}

\item{Uncertainties from ``tracking'' come from knowledge of the performance
of the tracking devices near the edges of the fiducial regions.  This performance
will depend largely on the tracking design and the collider environment near the 
beam. There will be abundant sources of tagged tracks with which to study 
tracking, and to define a ``good'' fiducial region.  We assume a factor five
improvement over the OPAL uncertainty.}

}

\end{itemize}

Combining the statistical and systematic errors in quadrature, the total error for the
measurement of the $e^+e^-\to \gamma\nu\bar{\nu}$ cross section at $\sqrt{s}=170$~GeV is
approximately 0.5\%. The corresponding allowed region is shown in  
Fig.~\ref{fig_LC}(left) in the $g_L\times g_R$-plane, 
assuming $N_{\nu}=3$ and that the measured 
central value coincides with the SM prediction for $(g_L,g_R)$, indicated by a star. 
As expected, the region is characterized by a ring in the $g_L\times g_R$-plane. 
However, since we have removed the kinematical region dominated by the radiative return 
to the $Z$-boson mass, the center of the ring is significantly displaced (to the left)
from the origin, while the radius of the ring is significantly larger than the one 
corresponding to the result obtained 
around the $Z$-boson mass (ring centered roughly around the origin). 
In the case of the SM, the two rings touch at a single point. 

The combination of the results obtained at $\sqrt{s}=m_Z$ and $\sqrt{s}=170$~GeV is
shown in the $(g_L^2+g_R^2)\times g_R/g_L$-plane in Fig.~\ref{fig_LC}(right).
This result is markedly more precise than the LEP+CHARM II result obtained earlier
(Fig.~\ref{fig_LEPcharm2}).
We would like to stress that the result depicted in Fig.~\ref{fig_LEPcharm2} is 
{\sl qualitatively}\/ different from the one depicted in Fig.~\ref{fig_LC}. 
In the former, we are combining very different data (obtained, for example, 
at very different center-of-mass energies), collected at completely different 
experiments. Consequently, assumptions are required in order to state that
the measurements are sensitive to the same physical parameters. In the
latter, we are comparing the same physical observable measured with the 
same detector, differing only by the center-of-mass energy.

\par
\begin{figure}[t]
\epsfig{figure=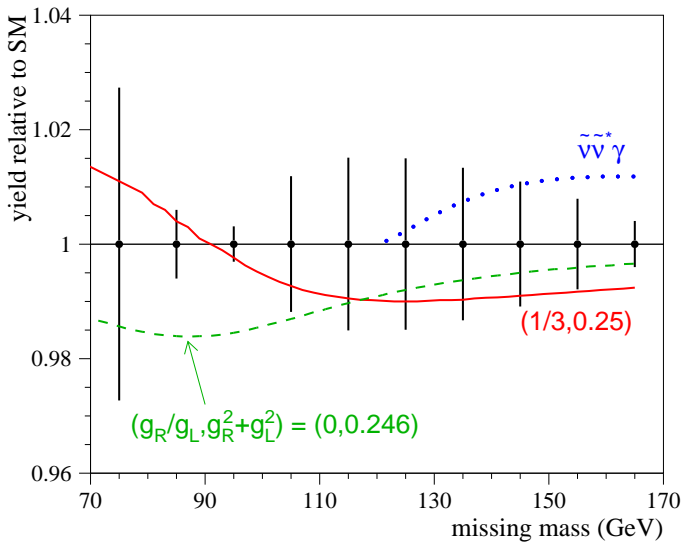,width=0.5\textwidth}%
\hspace{0.05\textwidth}%
\begin{minipage}[b]{0.45\textwidth}
\caption{Comparison of the shape of the missing mass distribution to
that expected in the SM.  The points with error bars indicate measurement
errors for $\sqrt{s} = 170$~GeV and $50~{\mathrm{fb}}^{-1}$.  The solid
curve shows the expected deviation when $g_R/g_L = 1/3$ and $g_R^2+g_L^2 = 0.25$,
while the dashed line shows $g_R = 0$ and $g_R^2+g_L^2 = 0.246$.  (In the SM,
$g_R = 0$ and $g_L^2 = 1/4$.)  The dotted line shows the contribution from
a single species of sneutrinos, with $M_{\tilde{\nu}} = 60$~GeV and no
contribution from $t$-channel chargino exchange.
\label{fig_meas_mm}}
\end{minipage}
\end{figure}

Thus far we have considered only the integrated cross section measured with 
a few kinematic cuts.  The relative contributions to 
${\rm d}\sigma_{\gamma\nu\bar{\nu}} / {\rm d}M_{\nu\bar{\nu}}$ 
depicted in Fig.~\ref{fig_comparison} depend on both $g_L$ and $g_R$.
The $ZZ$ term will change if either $g_L$ or $g_R$ varies, while the 
interference term, $WZ$, varies only with $g_L$.  The $WW$ term is
independent of both $g_L$ and $g_R$ under the assumption that the
charged weak interactions are the same as in the SM.  Consequently,
the shape of the missing mass distribution varies in a non-trivial way
as $g_L$ and/or $g_R$ deviate from their SM values.  We have estimated
the statistical errors for 10~GeV bins in the missing mass, and present
the result relative to the SM expectation in Fig.~\ref{fig_meas_mm}.
As illustration, we show the expected deviations for two sets of non-SM 
values for the $Z\nu\bar{\nu}$-couplings, both of which are allowed by current data.
In the first case, we take the SM value for $\Gamma_{\nu\bar{\nu}}$ but
allow $g_R$ to be one third of $g_L$.  The solid line shows the result:
no deviation at the $Z$-boson pole and a more or less constant reduction
in the cross section for $M_{\nu\bar{\nu}} > 105$~GeV.  In the second case,
we retain $g_R = 0$ as in the SM, but reduce $\Gamma_{\nu\bar{\nu}}$ by
less than~2\%. As indicated by the dashed line, a large deviation is
observed at the $Z$-boson mass, but it nearly disappears for high missing mass,
where the $WW$~term dominates. In this sense, a comparison of the bin 
$M_{\nu\bar{\nu}} \sim 90$~GeV to the bin $M_{\nu\bar{\nu}} \sim 165$~GeV
is tantamount to the NuTeV measurement of the $Z\nu\bar{\nu}$-coupling
suppression factor $\rho_0$.  Experimentally
this comparison would be exceptionally clean.

We conclude by commenting on the effect of relaxing the assumption that $N_{\nu}=3$. 
If one considers $N_{\nu}$ to be a free parameter
during the ``data'' analysis, no constraint on $g_R^2$, as defined in
Eq.~(\ref{define_gr_gl_Nnu}), 
can be obtained, while a ``measurement'' of $g_L$, to be performed in a way similar to our
measurement of $g_L$ from the LEP II data, can still be performed. This is easy to
understand. Because there is no $W$-boson exchange diagram for 
the $\nu_{\mu,\tau}\bar{\nu}_{\mu,\tau}$ final states, one could have redefined
\begin{equation}
N_{\nu}\left[(g_L^{\nu_{e}})^2+(g_R^{\nu_{e}})^2\right]\equiv (g_L^{\nu_{e}})^2+(g_{\rm others})^2,
\end{equation}
where $g_{\rm others}\equiv (N_{\nu}-1)(g_L^{\nu_{e}})^2+N_{\nu}(g_R^{\nu_{e}})^2$. It is easy
to see that, via $e^+e^-\to \gamma+$~invisible one is only sensitive to $g_L^{\nu_{e}}$ and
$(g_{\rm others})^2$, independent on whether $g_{\rm others}$ is the $Z$-boson 
coupling to the SM $\nu_{\mu,\tau}$, right-handed neutrinos, or other exotic invisible final states.

\setcounter{equation}{0}
\section{New physics contributions to the invisible $Z$-width}
\label{section_theory}

In the previous two sections, we have discussed a series of distinct experimentally
measurable quantities which are all closely related to $Z\nu\bar{\nu}$-couplings and
the number of active SM neutrinos. In particular, if the SM describes all the processes
discussed here, the direct and indirect measurements of the invisible $Z$-width should 
yield the same result (which may be translated into $N_{\nu}=3$), 
while measurements of $g_L^2+g_R^2$ and $g_L$ discussed in 
Sec.~\ref{section_away} should intersect at a single point: $g_L=+1/2$, $g_R=0$.
A statistically significant deviation of any of 
these measurements from SM predictions would signal that the SM is incomplete, and that
new physics is required in order to explain the values of these observables. 
In particular, it is possible that the direct and indirect measurements of the invisible $Z$-width
yield differing results, with the result that the two curves depicted in Fig.~\ref{fig_LC}
would intersect in either two or zero points.

Here, we will briefly discuss new physics mechanisms and/or models that will lead to 
physically observable effects in the measurements we discussed above. We first discuss several
mechanisms for modifying the invisible $Z$-width, concentrating on new physics that 
would modify the directly and indirectly measured invisible $Z$-width in distinct ways. Then, 
we argue whether one can construct a model with right-handed neutrino--$Z$-boson couplings,
and further discuss other ``applications'' of the $g_L\times g_R$ measurement discussed in 
Sec.~\ref{section_away} for constraining physics beyond the SM.

\subsection{Direct~$\times$~Indirect}

Several extensions of the SM will lead to an enhancement or suppression of the 
invisible $Z$-boson width with respect to SM expectations. Some of them modify 
the $Z$-boson decays in such a way that both the indirect and the direct
measurement of the invisible $Z$-width are modified in the same way ({\it i.e.,}\/
$\Gamma_{\rm inv}({\rm direct})=\Gamma_{\rm inv}({\rm indirect})\neq\Gamma_{\rm inv}^{SM}$).
For example, new decay modes of the $Z$-boson into invisible final states will 
enhance $\Gamma_{\rm inv}$ with respect to the SM prediction.
One example is the $Z$-boson decay into a pair of lightest 
neutralinos in R-parity conserving supersymmetry scenarios, $Z \to \tilde{\chi}^0_1 \,
\tilde{\chi}^0_1$, when the neutralinos are  predominantly
bino-like\footnote{Z-boson decays into neutralinos with a dominant wino or
higgsino component are already ruled out by present data.}.
While such contributions generically enhance the invisible $Z$-width,
different new physics effects may lead to
a reduction of the
magnitude of the $Z\nu\bar{\nu}$-couplings and hence suppress $\Gamma_{\rm inv}$.
This can be accomplished 
by assuming, for example, that the SM neutrinos mix slightly with sterile states. 
With the advent of the NuTeV anomaly
\cite{NuTeV}, which can be explained by reducing the $Z\nu\bar{\nu}$-couplings,
this option has recently received a significant amount of attention
(see, for example, \cite{theory_nutev,NuTeV_review}).  

Other effects can modify the indirectly measured value of the invisible
$Z$-width but not the one obtained directly. One mechanism that will lead to
such an effect is the following: assume that there is an exotic decay of the
$Z$-boson into final states with some charged and/or neutral particles. Such a decay will not
contribute to the direct measurement of the invisible  $Z$-width, as events
with detector activity other than a single photon are vetoed. On the other hand,
if these events fail the selection criteria for leptonic or hadronic $Z$-boson decays,
they will not contribute to $\Gamma_{\mathrm{vis}}$.  Since this new decay mode will
increase $\Gamma_{\rm tot}$ with respect to the SM prediction, 
$\Gamma_{\rm inv}({\rm indirect}) = \Gamma_{\rm tot}^{\rm SM} +
  \Gamma_{\rm tot}^{\rm new} - \Gamma_{\rm vis}^{\rm SM}
> \Gamma_{\rm inv}({\rm direct})$.
This might be the case, for example, if the $Z$-boson decays to a pair of
neutral particles which themselves decay some centimeters from the 
interaction point.
Another possibility is to introduce an effect that leads to 
$\Gamma_{\rm tot}^{\rm measured}\neq \Gamma_{\rm tot} =
\Gamma_{\rm vis}+\Gamma_{\rm inv}$. 
This will happen, for example, if another resonance is present ``on top'' of the
$Z$-pole \cite{Zprime_on_Z,sneu_res}.
Such a resonance will modify the line-shape of $e^+e^-\to f\bar{f}$
(which would no longer be a Breit-Wigner function) and lead to the extraction
of an effective total $Z$-boson width that differs from the ``real'' total
$Z$-boson width. On the other hand, if this new resonance does not decay into
invisible final states, no new contributions to the directly measured value of
$\Gamma_{\rm inv}$ will be present.  
One possibility of physics hidden by the $Z$-boson resonance is the $s$-channel exchange of
sneutrinos $\tilde{\nu}$ in R-parity violating supersymmetry
\cite{sneu_s_chan}. A sneutrino with mass close to the Z-boson mass,
$m_{\tilde{\nu}} \approx m_Z$, that primarily decays into $b\bar{b}$ pairs,
is not excluded by existing data from LEP and SLD and can lead to
deviations in the hadronic $Z$-boson line-shape parameters compared to the 
standard $Z$ line-shape parametrization
\cite{sneu_res}.

Finally, some new physics contributions can affect the directly measured value
of the  invisible $Z$-width but not the one obtained indirectly. The simplest
way of accomplishing this is to include new contributions to
$e^+e^-\to$~invisible that are not related to the $Z$-boson. For example, any
effective four-fermion interaction similar to
$\bar{e}\gamma^{\mu}e\bar{\nu}\gamma_{\mu}\nu$  contributes to
$e^+e^-\to\gamma\nu\bar{\nu}$ but does not contribute to $\Gamma_{\rm tot}$
extracted by the line-shape of the $Z$-boson resonance and hence to $\Gamma_{\rm
inv}(\rm indirect)$.
Such a four-fermion operator can be mediated, for example, 
by the exchange of extra neutral
gauge bosons, dubbed $Z'$-bosons. When the $Z'$-boson is relatively light, but
weakly coupled so that it forms a narrow resonance, the strongest experimental
bounds arise from the radiative return to the $Z'$-pole \cite{Zprime}.
It is, therefore, possible that effects of a $Z'$-boson are first discovered 
in the channel $\gamma\nu\bar{\nu}$. Note that, due to
possible interference effects between the $Z$-boson exchange and the new
effective interaction, the directly measured value of $\Gamma_{\rm inv}$ may be
suppressed or enhanced with respect to the SM prediction. Another option is to
consider the existence of anomalous $\gamma\nu\bar{\nu}$-couplings, 
which contribute to the electric and magnetic dipole moments and the
charge radii of the neutrinos \cite{Hirsch,gnunu_gen,nu_charad}.
Within the SM, effective $\gamma\nu\bar{\nu}$ interactions are generated at the
one-loop level, but have very small values \cite{nu_charad,gnunu_SM}.
However, loop effects from new physics can induce sizable
$\gamma\nu\bar{\nu}$ production rates \cite{Hirsch,gnunu_col}.
Such couplings contribute to $e^+e^-\to\gamma\nu\bar{\nu}$, but do not modify
measurements extracted from the $Z$-boson resonance.
Furthermore, other extensions to the SM introduce new invisible
particles, $Y$, that can be produced in $e^+e^- \to YY\gamma$. These include the Kaluza-Klein
gravitons of models with large extra dimensions \cite{extradim} and super-light
gravitinos in supersymmetry scenarios with gauge mediated supersymmetry 
breaking \cite{GMSBgrav}.

Note that in order to
enhance the experimental sensitivity to most of the mechanisms outlined in the
previous paragraph, one would profit from running at center-of-mass energies above the
$Z$-boson mass, in order to avoid the ``overwhelming presence'' of the
$Z$-boson resonance (see discussion in Sec. \ref{section_away}). 
On the other hand, some new physics effects lead to rare single-photon 
decays of the $Z$-boson, such as $Z\to\gamma\nu\nu$~\cite{rareZ}. In this case,
the measurement of the cross-section for
$e^+e^-\to\gamma\nu\bar{\nu}$ at the $Z$-pole yields valuable information.

Finally, we emphasize that more information can be obtained
by analyzing the missing mass distribution
${\rm d}\sigma_{\gamma\nu\bar{\nu}} / {\rm d}M_{\nu\bar{\nu}}$, 
as described in Sec.~\ref{section_away}B, and it may be possible to 
differentiate classes of new  physics contributions.
As illustrated in Fig.~\ref{fig_meas_mm}, excursions of $(g_L,g_R)$ from
the SM values would show up as distinguishable changes in the {\em shape} 
of the missing mass distribution.  Another possibility is the existence
of a new physics channel, for example, the production of sneutrino pairs
($e^+e^-\to \tilde{\nu}\tilde{\nu}^*\gamma$), as illustrated in 
Fig.~\ref{fig_meas_mm}, for the case $M_{\tilde{\nu}} = 60$~GeV. 
For this example, it is assumed that the sneutrinos 
are of the second or third generation, so that there is no contribution from
$t$-channel chargino exchange, and that the sneutrinos are stable or decay invisibly.
Clearly, the shape of the missing mass distribution allows the distinction of this 
contribution from any new physics effects that modify the properties of the $Z$-boson.
In a similar way, the contribution of an extra $Z'$-boson could be identified
by a resonance in the missing mass distribution, while the emission of
Kaluza-Klein gravitons in large extra dimensions
would yield a continuous background without threshold effects.

\subsection{Right-Handed Neutrino--$Z$-Boson Couplings?}

\begin{figure}[t]
\hspace{.05\textwidth}%
\epsfig{figure=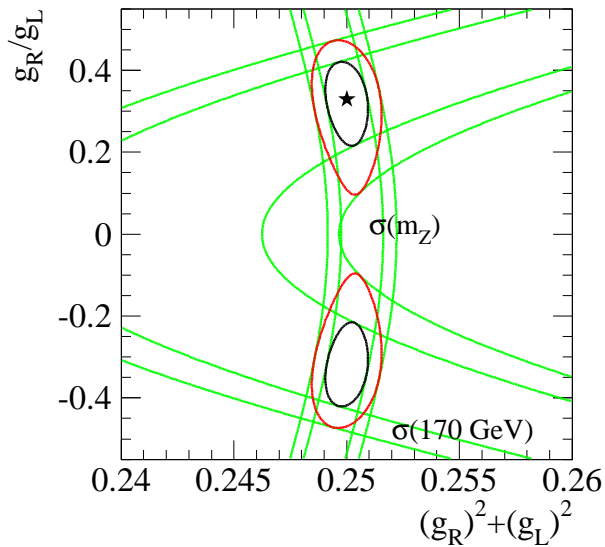,width=0.45\textwidth}%
\hspace{.05\textwidth}%
\begin{minipage}[b]{.45\textwidth}
\caption{Projected sensitivity of a linear collider to a non-zero right-handed
$Z\nu\bar{\nu}$ coupling $g_R = g_L/3$. As before, the analysis is based on
hypothetical measurements of the cross-section $e^+e^-\to\gamma\nu\bar{\nu}$ at
the $Z$-boson mass ($\sigma(m_Z)$) and at  $\sqrt{s}=170$~GeV ($\sigma(170~{\rm
GeV})$). The parameters of the underlying scenario are indicated by the star.
\label{fig_LCnosm}}
\end{minipage}
\end{figure}

In the SM, neutrinos couple only left-handedly to the $W$ and $Z$-bosons. 
This fact is a direct consequence of the $SU(2)_L\times U(1)_Y$-gauge symmetry 
structure of the SM, which fits almost all experimental data beautifully. 
On the other hand, we should not downplay the importance of directly verifying, 
experimentally, whether neutrino neutral currents are purely left-handed. 
Current data allow a right-handed $Z\nu\bar{\nu}$ coupling which is around 40\% as large 
as the left-handed one, while the LC measurement we propose could tighten the bound to 
about 30\%.
This should be contrasted with, say, our understanding of 
$Z\ell\bar{\ell}$-couplings and $W\ell\bar{\nu}$-couplings, which are known 
(in the worst case) at the few percent level.
In Fig.~\ref{fig_LCnosm} we show an example to
illustrate the sensitivity of a linear collider
to a non-zero right-handed $Z\nu\bar{\nu}$ coupling.
(The same example is also depicted in Fig.~\ref{fig_meas_mm}.)
Here, the right-handed coupling is chosen to be one third of the
left-handed coupling, which is allowed by current data,
while the value for the $Z$-width agrees with the SM prediction.
In the setup discussed in Sec.~\ref{section_away}B, it is possible to discriminate 
this scenario from the SM at more than the 
two sigma confidence level. There
remains, however, a twofold ambiguity, which is related to the fact that while the
sign of $g_L$ can be measured, the sign of $g_R$ remains undetermined.

It is interesting to probe whether there are new physics
models that lead to right-handed $Z\nu\bar{\nu}$-couplings. One example is to
consider the existence of a heavy $Z'$-boson that mixes slightly with the SM
$Z$-boson. In general, $Z$--$Z'$ mixing will lead to a shift in the SM $Z$-boson mass
and the SM $Z$-boson couplings to fermions. It is possible to choose $Z'$-boson couplings
to fermions such that: (i) the  left-handed $Z\nu\bar{\nu}$-coupling $g_L$ is
slightly reduced with respect to its SM value $g_L^{SM}$,  and (ii) a non-zero 
right-handed $Z\nu\bar{\nu}$-coupling $g_R$ is introduced.  If this is done in
such a way that $g_L^2+g_R^2\simeq (g_L^{SM})^2$, all current experimental
constraints can be safely evaded (see Sec.~\ref{section_away}A). 

We have constructed an explicit example, adding to the SM a 
$U(1)_{Z'}$ gauge symmetry under which leptons and right-handed neutrinos 
transform. 
In order to satisfy current experimental constraints and successfully introduce
a right-handed $Z\nu\bar{\nu}$-coupling we introduce two extra Higgs bosons. One, 
transforming nontrivially only under $U(1)_{Z'}$, is responsible for giving the $Z'$-boson
a mass. The other, which transforms under both $SU(2)_L$ and $U(1)_{Z'}$, is
responsible for inducing mixing between the SM $Z$-boson and the $Z'$-boson.

In the following, $Z$ and $Z'$ denote the mass eigenstates,
where the former corresponds to the physical $Z$-boson that has been observed
at LEP and other colliders, while the eigenstates of the electroweak gauge group and the
extra gauge group are given by $Z_1$ and $Z_2$, respectively. Using this
language, we can say that the interference
between $Z_1$ and $Z_2$ can lead to a reduction of the $Z\nu_L\bar{\nu}_L$-coupling, 
while a $Z\nu_R\bar{\nu}_R$-coupling is introduced by the $Z_2$
admixture in the $Z$-boson. Several constraints have to be taken into account, however.
First of all, because the $Z'$ couples to charged leptons, it is currently constrained to be
very heavy, $m_{Z'}\gtrsim 900$~GeV \cite{PDG}. Second, in order to induce a relatively
large right-handed $Z\nu\bar{\nu}$-coupling, we are required to have either
sizable $Z$--$Z'$-mixing
and/or a very large $U(1)_{Z'}$ coupling. Large $Z$--$Z'$-mixing will imply in a significant 
shift of the $Z$-boson mass from SM expectations, especially because the $Z'$ is constrained
to be very heavy. Third, one should keep in mind that not only are the $Z\nu\bar{\nu}$-couplings
modified, but also the $Z\ell\bar{\ell}$-couplings. Taking all of these
constraints into account, we are able to find an
``existence-proof'' example. We choose a $U(1)_{Z'}$ coupling $g'=3.5$, and set the charges
of the left-handed leptons and the right-handed charged leptons to $+1/2$. We also set the
charges of the right-handed neutrinos to $+5$.\footnote{Here we did not consider anomaly cancellation. 
We assume that this can be accomplished for example by adding heavy fermions to the theory.}  
We further set $m_{Z'}=1$~TeV and the $Z$--$Z'$-mixing angle to $\sin^2\theta_{ZZ'}=1.6\times 10^{-5}$. 
Under these conditions, the $Z\nu_L\bar{\nu}_L$ coupling $g_L$ is reduced by 3.7\%, while a right-handed
coupling is generated: $|g_R/g_L|=0.19$. 
We further verified that the above mentioned shifts
to the $Z\ell\bar{\ell}$-couplings and the $Z$-boson mass are allowed by the data, and find a fit
to $\Gamma_{\rm tot}$, $\sin^2\theta_{W,\rm eff}^{\ell}$, $\sin^2\theta_{W,\rm eff}^{\rm had}$, 
$R_{\ell}$, and $\sigma_h^0$ that is as satisfactory as the SM fit.

We stress that this model is not intended to provide a realistic description of
nature, but only to prove the possibility of ``large'' right-handed
$Z\nu\bar{\nu}$ couplings that are not excluded by existing data.  While the
specific value, $|g_R/g_L|=0.19$, we obtained above would still not be
detectable at a linear collider, we emphasize that there are other, more
complicated, $Z'$ models  that induce larger $Z\nu_R\bar{\nu}_R$-couplings. 
For example, the $Z\nu_L\bar{\nu}_L$-coupling could be further reduced through
the addition of higher dimensional operators to the model, thus allowing an
increase of the value of the $Z\nu_R\bar{\nu}_R$-coupling to $|g_R/g_L|>0.3$,
which can be distinguished from $g_R=0$ at more than the two sigma confidence
level at a linear collider, as depicted in Fig.~\ref{fig_LCnosm}.

More generally, as briefly alluded to in Sec.~\ref{section_away}, the $g_R$ need not be
a right-handed coupling of the neutrino to the $Z$-boson, but can be interpreted
as any coupling of the $Z$-boson to exotic invisible final states. Within the SM, 
the magnitude of these couplings is best constrained by measuring the invisible $Z$-width 
at the $Z$-boson mass, as discussed above. However, by performing the analyses discussed 
in Sec.~\ref{section_away}, one is capable of separating the $Z$-boson coupling to active 
neutrinos ($g_L$) from the $Z$-coupling to exotic invisible matter. These analyses allow 
one to identify models where the $g_L$ coupling is smaller than usual, but is somehow 
``compensated'' by the exotic contribution. 
If this is the case, the invisible $Z$-width measurements at the $Z$-boson mass do not 
register a discrepancy with respect to the SM, while the measured value of $(g_L,g_R)$ 
deviates from $(1/2,0)$ (this is exactly what happens in the $U(1)_{Z'}$ model 
spelled out above). 

\setcounter{equation}{0}
\section{Summary and Conclusions}
\label{section_conclusions}

We have discussed how measurements at $e^+e^-$ colliders provide information
on $Z\nu\bar{\nu}$-couplings in distinct ways, allowing the exploration of
new physics contributions to the left-handed and right-handed neutrino 
couplings to the $Z$-boson.  
The {\sl indirect} measurement of $\Gamma_{\rm inv}$ is obtained by 
subtracting the visible partial width from the total width of the
$Z$-boson resonance. It provides a tight constraint on $g_L^2+g_R^2$.
The {\sl direct} measurement of $\Gamma_{\rm inv}$ comes
from measuring the cross section for $e^+e^-\rightarrow
\gamma\nu\bar{\nu}$.  When this done at center-of-mass energies
around the $Z$-boson pole mass, it is again possible to constrain the combination
$g_L^2+g_R^2$. At higher center-of-mass energies, however, 
good sensitivity to $g_L$ is obtained from the interference of the $Z$-boson
and $W$-boson exchange amplitudes.  

We examined published data from LEP in order to constrain
$g_L$ and $g_R$.  The strongest constraint by far comes from the 
indirect value of $\Gamma_{\rm inv}$.  We have also analyzed
data taken at energies above the $Z$-pole (LEP~II) in order to extract $g_L$.
Although the result obtained is not particularly precise, in part because only a fraction of the 
collected data have been analyzed by the LEP collaborations, it establishes the sign 
of $g_L$ (positive) at the $2\sigma$ confidence level. 
If all existing LEP data were analyzed with the requirement that
the missing mass be greater than $100$~GeV, we estimate that this analysis would 
establish $g_L$ with an uncertainty $\delta(g_L) = 0.05$.

Important constraints on $|g_L|$ also come from measurements of elastic
neutrino--electron scattering by CHARM~II and LAMPF.  Their results agree
well with the ones provided by LEP, and with SM predictions.  
Combining LEP and CHARM II data, the value
of the left-handed $Z\nu\bar{\nu}$-coupling is constrained to be 
$0.45 \lesssim g_L \lesssim 0.5$, while
the right-handed neutrino couplings to the $Z$-boson are mildly bounded to be 
$|g_R|\lesssim 0.2$ at the $2\sigma$ level.
The NuTeV data can also be used to determine  
$Z\nu\bar{\nu}$-couplings, in which case a very tight constraint on $|g_L|$ 
would be obtained. This interpretation, however, rests on several assumptions 
which are not universally accepted.  In our opinion, the elastic
neutrino--electron scattering results are cleaner, and we speculate
that a new experiment using existing or future neutrino beams should
improve substantially the precision with which $|g_L|$ is measured.

A future $e^+e^-$ linear collider could run at center-of-mass energies near 
the $Z$-boson pole mass in the so-called ``Giga-Z'' option.  The integrated 
luminosities are expected to be much larger than those recorded at LEP, on the order 
of $50~{\mathrm{fb}}^{-1}$. We have estimated how new data taken around the 
$Z$-boson pole mass and at $\sqrt{s} = 170$~GeV would improve the constraints 
already obtained from LEP data. 
We find that the quality of the indirect measurement of $\Gamma_{\rm inv}$ 
would be only modestly improved, while the direct measurement would be performed with
greatly improved precision. We estimate that at a linear collider the
precision with which the $Z$-boson width could be directly measured would be
comparable to the precision of the indirect measurement. 
Since both measurements are potentially sensitive to different new physics, a 
high precision in both is very desirable. 

In particular, measurements of $e^+e^-\to\gamma+$~invisible 
at $\sqrt{s} = m_Z$ and $\sqrt{s} = 170$~GeV would be sensitive to 
values of $g_R$ on the order of $g_L/3$.
The data samples should be large enough to warrant a measurement of the
cross section as a function of missing mass.  Deviations of this
differential cross section from the SM prediction would indicate 
whether $g_L$, $g_R$, or both differ from the SM value.
The fact that one experiment running at different energies
can simultaneously constrain both $g_L$ and $g_R$ makes the
Giga-Z option particularly attractive in this context.
The comparison of these two energy regions
will provide information analogous to the NuTeV determination of the
$Z\nu\bar{\nu}$-couplings.

Finally, we have sketched a variety of new physics scenarios that
will impact the measurements of the indirect and direct 
invisible $Z$-width differently, many of which have been already considered in the 
literature in various contexts. In summary, by performing both the 
direct and indirect measurements of the invisible $Z$-width with similar 
precision, one should be able to distinguish between several different 
new physics mechanisms, depending on whether the measurements agree or disagree
with SM predictions, or whether the two distinct measurements of the invisible
$Z$-width agree or disagree with each other. We have also discussed the importance
of measuring ``$g_R$'' as far as constraining new physics. The main reason for this
is the fact that while we generally referred to right-handed $Z\nu\bar{\nu}$-couplings,
other couplings of the $Z$-boson to exotic, invisible final states are probed in exactly the
same way. 

We conclude by restating the interesting fact that the current data, 
in particular the NuTeV anomaly and the LEP measurement of 
$\Gamma_{\rm inv}$, hint at a non-standard  $Z\nu\bar{\nu}$-couplings. 
Only future experiments can elucidate this issue.

\section*{Acknowledgments}

AdG would like to thank the KITP in Santa Barbara and the ICTP in Trieste for 
their hospitality during, respectively, the early stages and the final stages 
of this work.
MC would like to thank the Aspen Center for Physics for its hospitality 
during the completion of this paper.
The work of MC, AdG and AF is supported by the US Department of Energy 
Contract DE-AC02-76CHO3000.
The work of MS is supported by US DoE Contract DE-FG02-91ER40684,
and by the Illinois Consortium for
Accelerator Research, agreement number~228-1001.
\pagebreak[0]

\end{document}